\documentclass{emulateapj}
\usepackage{apjfonts}

\newcommand{\beq}{\begin{equation}}
\newcommand{\eeq}{\end{equation}}

\def\farcm{\hbox{$.\mkern-4mu^\prime$}}
\def\farcs{\hbox{$.\!\!^{\prime\prime}$}}
\def\farchs{\hbox{$.\!\!^{\rm s}$}}

\def\arcmin{\hbox{$^\prime$}}
\def\arcsec{\hbox{$^{\prime\prime}$}}
\def\solar{\mbox{$_{\normalsize\odot}$}}

\def\deg{\hbox{$^\circ$}}

\newcommand{\AmS}{{\protect\the\textfont2
  A\kern-.1667em\lower.5ex\hbox{M}\kern-.125emS}}
\newcommand{\lsim}{\ \raise
-2.truept\hbox{\rlap{\hbox{$\sim$}}\raise5.truept\hbox{$<$}\ }}
\newcommand{\gsim}{\ \raise
-2.truept\hbox{\rlap{\hbox{$\sim$}}\raise5.truept\hbox{$>$}\ }}
\newcommand{\simsim}{\ \raise
-2.truept\hbox{\rlap{\hbox{$\sim$}}\raise5.truept\hbox{$\sim$}\ }}

\slugcomment{Accepted for Publication in the Astrophysical Journal}

\righthead{Stellar Stratification
in Young Star Clusters in the Large Magellanic Cloud}
\lefthead{Gouliermis, et al.}

\shorttitle{Stellar Stratification
in Young Star Clusters in the Large Magellanic Cloud}
\shortauthors{Gouliermis, et al.}

\begin{document}
 
\title{Assessment of Stellar Stratification in Three Young Star Clusters in the Large Magellanic Cloud.}



\author{Dimitrios A. Gouliermis}
\affil{Max Planck Institute for Astronomy, K\"onigstuhl 17, 69117 Heidelberg, Germany}
\email{dgoulier@mpia-hd.mpg.de}
\author{Dougal Mackey}
\affil{Institute for Astronomy, University of Edinburgh, Royal Observatory,\\ 
Blackford Hill, Edinburgh, EH9 3HJ, UK}
\email{dmy@roe.ac.uk}
\author{Yu Xin}
\affil{Argelander-Institut f\"{u}r Astronomie, Rheinische Friedrich-Wilhelms-Universit\"{a}t Bonn,\\
Auf dem H\"{u}gel 71, 53121 Bonn, Germany}
\email{yxin@astro.uni-bonn.de}
\and
\author{Boyke Rochau}
\affil{Max Planck Institute for Astronomy, K\"onigstuhl 17, 69117 Heidelberg, Germany}
\email{rochau@mpia-hd.mpg.de}

%

\begin{abstract} 
We present a comprehensive study of stellar stratification in young star clusters in 
the Large Magellanic Cloud (LMC). We apply our recently developed {\sl effective 
radius method} for the assessment of stellar stratification on imaging data obtained  
with  the Advanced Camera for Surveys of three young LMC clusters to characterize 
the phenomenon and develop a comparative scheme for its assessment in such 
clusters. The clusters of our sample, NGC~1983, NGC~2002 and NGC~2010, are 
selected on the basis of their youthfulness, and their variety in appearance, structure, 
stellar content, and surrounding stellar ambient. Our photometry is complete for magnitudes 
down to $m_{\rm 814} \simeq 23$~mag, allowing the calculation of the structural 
parameters of the clusters, the estimation of their ages and the determination of their 
stellar content. Our study  shows that each cluster in our sample demonstrates stellar 
stratification in a quite different manner and at different degree from the others. Specifically, 
NGC~1983 shows to be {\sl partially} segregated with the effective radius increasing with 
fainter magnitudes only for the faintest stars of the cluster. Our method on NGC~2002 
provides evidence of {\sl strong} stellar stratification for both bright and faint stars; 
the cluster demonstrates the phenomenon with the highest degree in the sample. Finally, NGC~2010 is {\sl not 
segregated}, as its bright stellar content is not centrally concentrated, the relation 
of effective radius to magnitude for stars of intermediate brightness is rather flat, and 
we find no evidence of stratification for its faintest stars. 
For the parameterization of the phenomenon of stellar stratification and its quantitative 
comparison among these clusters, we propose the slope derived from the change 
in the effective radius over the corresponding magnitude range as indicative parameter of 
the {\sl degree of stratification} in the clusters. A positive value of this slope
indicates mass segregation in the cluster, while a negative or zero value signifies the lack of the 
phenomenon.

\end{abstract}


\section{Introduction}

{\sl Stellar stratification} is a characteristic phenomenon of star clusters, 
which has been well documented for more than 50 years. It is 
directly linked to the dynamical evolution of star clusters 
\citep{chandrasekhar42, spitzer87},  and specifically to the central 
concentration of the massive stars as the cluster relaxes dynamically, 
a phenomenon known as {\sl mass segregation}
\citep[for detailed reviews see][]{lightman78,meylan97}. This behavior 
in massive Galactic globular clusters (GGCs) is well understood and the observed 
mass segregation is explained as a result of their dynamical relaxation in a 
timescale, $t_{\rm relax}$, much longer than their ages, 
$\tau$, as derived from stellar evolutionary models. This {\sl dynamical} mass 
segregation, thus, forces more massive stars to sink inwards to the center 
of a cluster through weak two-body interactions.

However, segregation of massive stars is observed also in young star clusters, such as 
the $\sim$~30~Myr-old cluster NGC~330 in the Small Magellanic Cloud 
\citep{sirianni02}, or even the $\sim$~1~Myr-old Galactic starburst NGC~3603
\citep{stolte06}. Such systems, supposedly not sufficiently old to be dynamically relaxed,
cannot be included in the ``traditional'' picture of dynamical mass 
segregation as it occurs in GGCs. For these cases the phenomenon has 
been -- rather tentatively -- characterized as {\sl primordial} mass 
segregation, based on the suggestion that this stellar stratification 
is due to the formation of the stars in, or near, the central part of 
the clusters rather than being dynamically driven
   
Mass segregation at early stages of clustered star formation is indeed 
predicted by theoretical studies, which suggest that the positions of 
massive stars for rich young clusters of $\tau << t_{\rm relax}$ cannot 
be the result of dynamical evolution \citep{bonnell98}. Moreover, the 
appearance of gas in such clusters reduces further the efficiency of 
any dynamically driven mass segregation. According to theory, massive 
protostars can be formed {\sl at the central part} of the protocluster 
through cohesive collisions and dissipative merging of cloudlets, which 
lead to extensive mass segregation and energy equipartition among
them \citep{murray96}, or through accretion from a distributed 
gaseous component in the cluster, which leads to mass segregation
since stars located near the gravitational center of the cluster 
benefit from the attraction of the full potential and accrete at 
higher rates than other stars \citep{bonnell06}. 

Both of the aforementioned massive star formation 
mechanisms predict primordial mass segregation through early dynamical 
processes at the central part of the protocluster. The significant 
difference of these processes from the subsequent energy equipartition 
of the stars in the formed cluster is that the latter, and consequently 
dynamical mass segregation, is independent of the initial cluster 
conditions, while the former, and therefore primordial mass segregation, 
is directly connected to these conditions. In order to 
achieve a complete understanding of the phenomenon of primordial mass 
segregation, many related studies are focused on clusters at very early 
stages of their formation, including a variety of types of stellar systems 
from loose stellar associations and open clusters to compact massive starbursts.

The Magellanic Clouds (MCs), being extremely rich in young star clusters 
\citep[e.g.,][]{bica95, bica99}, offer an outstanding sample of intermediate 
type of stellar systems, specifically compact young star clusters, 
which are not available in the Galaxy. In such clusters primordial mass 
segregation can be sufficiently observed from ground-based observations 
\citep[e.g.,][]{kontizas98, santiago01, kerber02, kumar08}, and they are also 
suitable for more detailed studies of this phenomenon from space with the 
high-resolving efficiency of the {\sl Hubble Space Telescope} \citep[e.g.][]{fischer98, 
degrijs02, sirianni02, gouliermis04, kerber06}. Such studies contribute significantly
to the debate about the phenomenon of primordial mass segregation and its origin 
which which is still ongoing \citep[see, e.g.,][]{kisslerpatig07}. It is interesting to note 
that while the existence of the phenomenon itself is being questioned 
\citep{ascenso09}, new methods are still being designed for its comprehensive 
identification and quantification \citep[e.g.,][]{allison09}.

In the first part of our study of stellar stratification we developed and tested a new 
robust diagnostic method aimed at establishing the presence (or not) of this phenomenon 
in a cluster  \citep[][from here on Paper~I]{gouliermis09}. 
This method is based on the calculation of the mean-square radius, the so-called 
{\sl effective radius}, of the stars in a cluster for different magnitude ranges, 
and the investigation of the dependence of the effective radii of each stellar 
species on magnitude as indication of stellar stratification in the cluster. 

\begin{figure}[t!]
\centerline{\includegraphics[clip=true,width=0.45\textwidth]{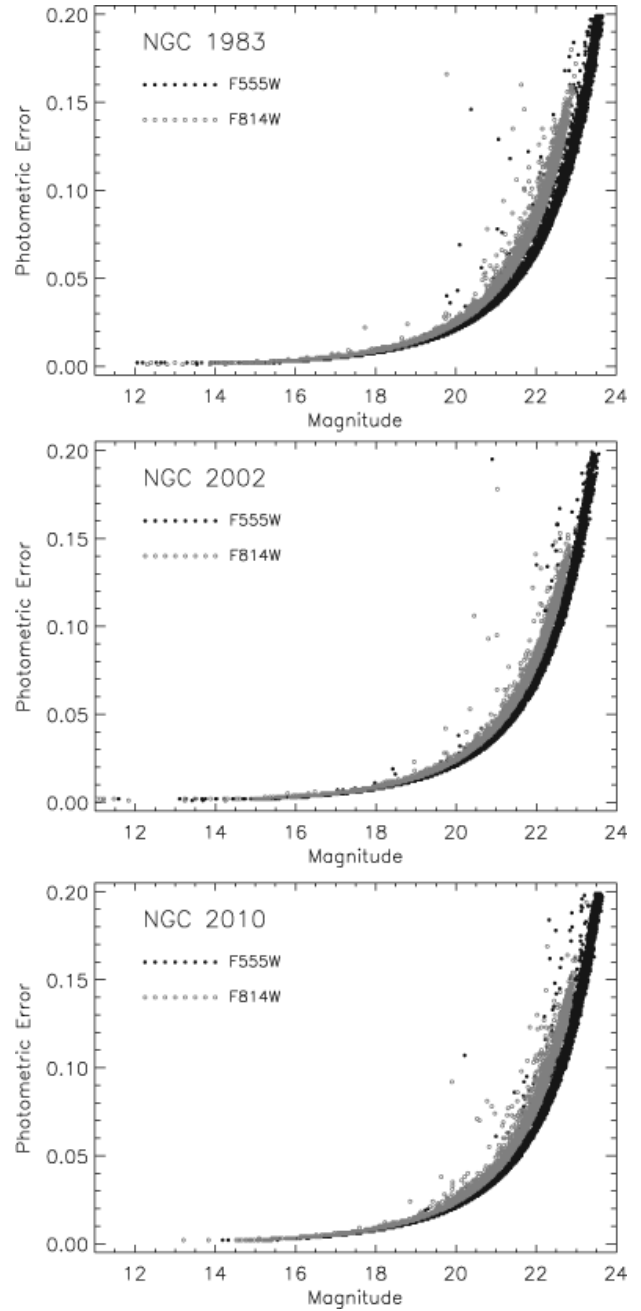}}
\caption{Distribution of the uncertainties of our photometry for all stars detected in 
both the F555W and F814W bands in the areas of the clusters of our sample.
Uncertainties in F555W are plotted with filled black and in F814W in open grey 
circular symbols. 
\label{f:phterr}}
\end{figure}

\subsection{The Effective Radius Method \label{s:method}}

Stellar stratification affects the presently observable physical properties 
of the star cluster, and thus diagnostic tools for the identification of the 
phenomenon are based on the detection of mass- or brightness-dependent changes 
of such properties. Previously established methods involve (i) the study of the 
projected stellar density distribution of stars of different magnitudes; here, 
the differences in the exponents of the slopes fitted are seen as proof of 
mass segregation \citep[e.g.,][]{subramaniam93}, (ii) the 
radial dependence of the Luminosity or Mass Function (LF, MF) slope; a gradient 
of this slope outwards from the center of the cluster indicates stratification  
\citep[e.g.,][]{king95}, and (iii) the variance of the core radius of the cluster 
as estimated for stars in specific magnitude (or mass) groups; a trend of this 
variance with brightness (or mass) signifies stratification \citep[e.g.,][]{brandl96}.
While these methods can provide useful information about the segregation in a  
cluster, all of them are sensitive to model assumptions, since they strongly depend 
on data fitting for parametrizing stellar stratification; i.e., functional
expressions for the surface density profiles, or the MFs and LFs are required for the 
derivation of their slopes and core radii. 

Our method, developed in Paper~I for assessment of stellar stratification in star 
clusters, is based on the notion of the dynamically stable {\sl Spitzer radius} of a 
star cluster, defined as the mean-square distance of the stars from the 
center of the cluster \citep[e.g.,][]{spitzer58}. The observable counterpart
of this radius, the effective radius, is given by the expression: 
\begin{equation} \displaystyle{r_{\rm eff} = \sqrt{\frac{\displaystyle{\Sigma_{i=1}^{N}} 
r_{i}^{2}}{N} }}, \label{eq:reff}\end{equation} where $r_{i}$ is the projected radial distance 
of the $i$th stellar member of the cluster in a specific brightness range and
$N$ the corresponding total number of stars in the same brightness range. 
Different calculations of the effective radius are performed, each for stars 
in different brightness ranges that cover the whole observed stellar luminosity 
function of the cluster. The simple application of this method can yield direct
information about the spatial extent (radial distribution) of stellar
groups in different magnitude (mass) ranges. In a star cluster, where stratification 
occurs, the segregated brighter stars are expected to be more centrally 
concentrated and the corresponding effective radius should be 
shorter than that of the non-segregated fainter stars. As a consequence, 
stellar stratification can be observed from the dependence of the effective 
radius of stars in specific magnitude (mass) ranges on the corresponding mean 
magnitude (mass). This method is sufficiently described and tested in Paper~I, 
where we show that it performs efficiently in the detection of stellar 
stratification, provided that the incompleteness of the photometry has been 
accurately measured and the contamination by the field population has been 
thoroughly removed. Our diagnosis method is also independent of any model or 
theoretical prediction, in contrast to those used thus far for the detection 
of mass segregation. 


\begin{deluxetable*}{ccccccc}
\label{t:obs}
\tablecolumns{8}
\tablewidth{0pc}
\tablecaption{ACS/WFC observations of the clusters of our sample (HST Program 9891). \label{t:obs} }
\tablehead{
\colhead{Cluster} & 
\colhead{R.A.} &
\colhead{DEC} &
\colhead{Filter} &
\colhead{Dataset} &
\colhead{Exposure} &
\colhead{Date} \\
\colhead{} & 
\colhead{(J2000.0)} &
\colhead{(J2000.0)} &
\colhead{} &
\colhead{filename} &
\colhead{time (s)} &
\colhead{} 
} 
\startdata
NGC~1983 &  05$^{\rm h}$27$^{\rm m}$44\farchs92& $-$68\deg59\arcmin07\farcs0& F555W & j8ne68req & 20 & October 7, 2003\\
                     &  &  & F814W & j8ne68riq & 20 & October 7, 2003\\
NGC~2002 &  05$^{\rm h}$30$^{\rm m}$20\farchs80& $-$66\deg55\arcmin02\farcs3& F555W & j8ne70b5q & 20 & August 23, 2003\\
                     &  &  & F814W & j8ne70b7s & 20 & August 23, 2003\\
NGC~2010 & 05$^{\rm h}$30$^{\rm m}$34\farchs89& $-$70\deg49\arcmin08\farcs3& F555W & j8ne71r5q & 20 & October 7, 2003\\
                     &  &  & F814W & j8ne71r9q & 20 & October 7, 2003\\
\enddata
\end{deluxetable*}

In this second part of our study of stellar stratification we apply the effective 
radius method on deep imaging data obtained with the Wide-Field Channel (WFC) of the 
{\sl Advanced Camera for Surveys} (ACS) onboard the Hubble Space Telescope (HST) 
of three young star clusters in the Large Magellanic Cloud (LMC) for the investigation of 
primordial mass segregation in them. Young clusters in the Magellanic Clouds are 
quite different to each other, so that their study requires the detailed treatment of each cluster
individually (see, e.g., \S\S~\ref{s:dynamics}, \ref{s:starcont}), and therefore a statistical 
investigation of the phenomenon of mass segregation in a large sample of such clusters 
is not yet meaningful.  The purpose of the present study is the detailed investigation with 
the use of a consistent method of individual selected clusters, 
in order to establish a comprehensive comparative framework for the assessment 
of mass segregation in young LMC clusters. The objects 
of interest, clusters NGC~1983, NGC~2002 and NGC~2010, are selected among others
also observed with ACS due to their youthfulness and the variety in  their characteristics. 
This selection is made in order to include in this comparative study clusters, which are
very  different from each other.  In \S~\ref{s:clussampl} we describe the considered sample of 
 LMC clusters and the corresponding data sets, as well as their reduction and photometry. 
We study the dynamical status of the clusters and discuss their structural parameters
in \S~\ref{s:dynamics}, and in \S~\ref{s:starcont} we present the observed stellar populations, 
the decontamination of the stellar samples from the contribution of the local background 
field of the LMC and the stellar content of the clusters. The application of the diagnostic 
method is performed for all three clusters in \S~\ref{s:methappl}, where we also develop 
a comparative scheme for the quantification of stellar stratification among different clusters, 
and we discuss our results. We conclude on them in \S~\ref{s:concl}.

\begin{figure}[t!]
\centerline{\includegraphics[clip=true,width=0.45\textwidth]{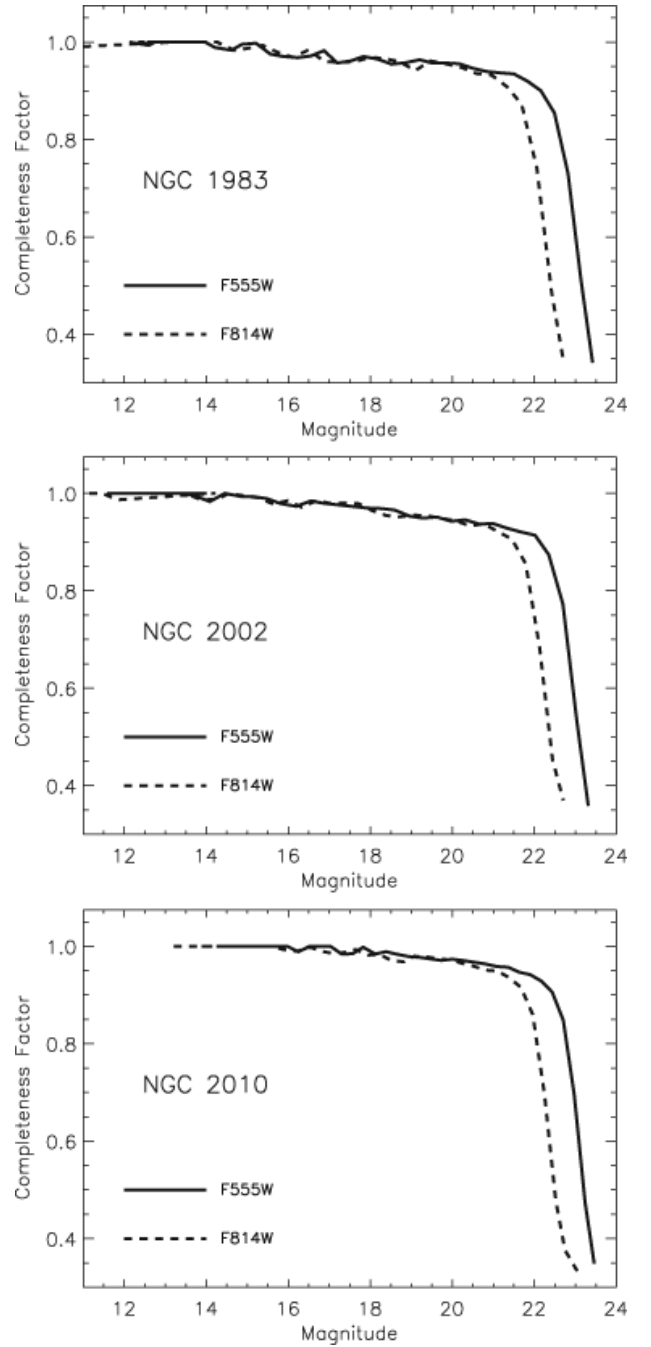}}
\caption{Completeness functions for NGC~1983 (top), NGC~2002 (middle), and 
NGC~2010 (bottom) for both filters F555W (solid line) and  F814W (dashed line).
Average completeness functions over the whole observed fields of view are shown. 
It should be noted that completeness is a function of cluster-centric distance, as stars 
closer to the center of a cluster suffer from high crowding than those farther away. 
\label{f:compl}}
\end{figure}

\section{The Star Cluster Sample}\label{s:clussampl}

The present study deals with stellar stratification in rich LMC star 
clusters. Such clusters are known to differ from each other in terms 
of, e.g., their structural parameters, ages, luminosities and masses 
\citep{mackey03}. Since we focus on the phenomenon of 
primordial mass segregation we limit our sample to young LMC clusters, 
and in order to use the most complete available data we select clusters 
observed with ACS/WFC. The advantages 
introduced to crowded-field photometry of MCs star clusters by the use of 
ACS, and specifically its Wide-Field Channel, have been documented extensively 
by studies based on such observations \citep[e.g.,][]{mackey06, rochau07, xin08}. 
Consequently, we select three young LMC clusters, NGC~1983, NGC~2002 and 
NGC~2010, observed with ACS/WFC. These clusters are different 
from each other, as far as their appearance, structure and stellar content is concerned. 
Considering that we also wish to investigate the identification and quantification 
of the phenomenon of stratification in various cluster-hosting environments, 
we based the selection of these specific clusters on the on the 
differences between their local surrounding field star populations. 
We discuss in detail the differences of the clusters themselves in their structure and 
dynamical behavior in \S~\ref{s:dynamics}, and in their stellar content in 
\S~\ref{s:starcont}. The decontamination of the true cluster populations 
from the contribution of the general background field of the LMC, which 
determines the actual stellar content of the clusters and the stellar sample 
to be used in our method for assessment of stellar stratification in the clusters 
is performed in \S~\ref{ss:fldsub}.  

\subsection{Data Reduction and Photometry}\label{ss:phot}

Our observations come from HST program 9891,
a snapshot survey of $\sim 50$ rich star clusters in the Magellanic 
Clouds. Imaging was obtained in Cycle 12 using ACS/WFC. 
As snapshot targets, 
each cluster was observed for one orbit only, resulting in a single 
exposure through the F555W filter and another through the F814W filter. 
Details of the observations for the clusters considered in this paper
may be found in Table \ref{t:obs}.

The ACS WFC consists of a mosaic of two $4096\times2048$~pixel CCDs 
covering an area of approximately $202\arcsec \times 202\arcsec$ at a plate
scale of $0.05$ arcsec per pixel. The two CCDs are separated by a gap
roughly $50$ pixels wide. The core of each cluster was placed at the
center of WFC chip 1. This ensured that the inter-chip gap did not
interfere with the main part of the cluster, and meant that each image
reached a maximum radius of approximately $\sim 150\arcsec$ from
the cluster center. A small dither was made between the two exposures
of a given target, to help facilitate the removal of cosmic rays and
hot pixels; however with only two images per cluster it was not possible
to eliminate the gap in spatial coverage due to the inter-chip separation.

The data products produced by the standard STScI reduction pipeline,
which we retrieved from the public archive, have had bias and 
dark-current frames subtracted and are divided by a flat-field image. 
In addition, known hot-pixels and other defects are masked, and the 
photometric keywords in the image headers are calculated. We also 
obtained distortion-corrected (drizzled) images from the archive, 
produced using the {\sc pyraf} task {\sc multidrizzle}.

We used the {\sc dolphot} photometry software \citep[e.g.,][]{dolphin00},
specifically the ACS module\footnote{The ACS 
module of {\sc dolphot} is an adaptation of the photometry package 
{\sc HSTphot} \citep{dolphin00}. The package can be retrieved from 
{\tt http://purcell.as.arizona.edu/dolphot/}.}, to photometer the flatfielded (but not 
drizzled) F555W and F814W images. {\sc dolphot} performs point-spread 
function (PSF) fitting using PSFs especially tailored to the ACS camera. 
Before performing the photometry, we first prepared the images using the 
{\sc dolphot} packages {\sc acsmask} and {\sc splitgroups}. Respectively, 
these two packages apply the image defect mask and then split the 
multi-image STScI FITS files into a single FITS file per chip. We then 
used the main {\sc dolphot} routine to simultaneously make photometric
measurements on the pre-processed images, relative to the coordinate 
system of the drizzled F814W image. We chose to fit the sky locally 
around each detected source (important due to the crowded nature of the 
targets), and keep only objects with a signal greater than $10$ times 
the standard deviation of the background. The output photometry from 
{\sc dolphot} is on the calibrated VEGAMAG scale of \cite{sirianni05}, 
and corrected for charge-transfer efficiency (CTE) degradation.

To obtain a clean list of stellar detections with high quality photometry,
we applied a filter employing the sharpness and ``crowding'' parameters 
calculated by {\sc dolphot}. The sharpness is a measure of the broadness 
of a detected object relative to the PSF -- for a perfectly-fit star this 
parameter is zero, while it is negative for an object which is too sharp 
(perhaps a cosmic-ray) and positive for an object which is too broad 
(e.g., a background galaxy). The crowding parameter measures how much 
brighter a detected object would have been measured had nearby objects 
not been fit simultaneously. We selected only objects with 
$-0.15 \leq {\rm sharpness} \leq 0.15$ in both frames, and 
${\rm crowding} \leq 0.25$ mag in both frames. We also only kept objects 
classified by {\sc dolphot} as good stars\footnote{Which have object type 
$1$, as opposed to elongated or extended objects which have object types 
$> 1$.} with formal errors from the PSF fitting less than $0.2$ mag in 
both F555W and F814W. In Figure~\ref{f:phterr} typical uncertainties in the 
derived magnitudes of the stars detected in all three clusters are shown for 
both F555W and F814W filters.

\begin{figure*}[t!]
\centerline{\includegraphics[clip=true,width=0.725\textwidth]{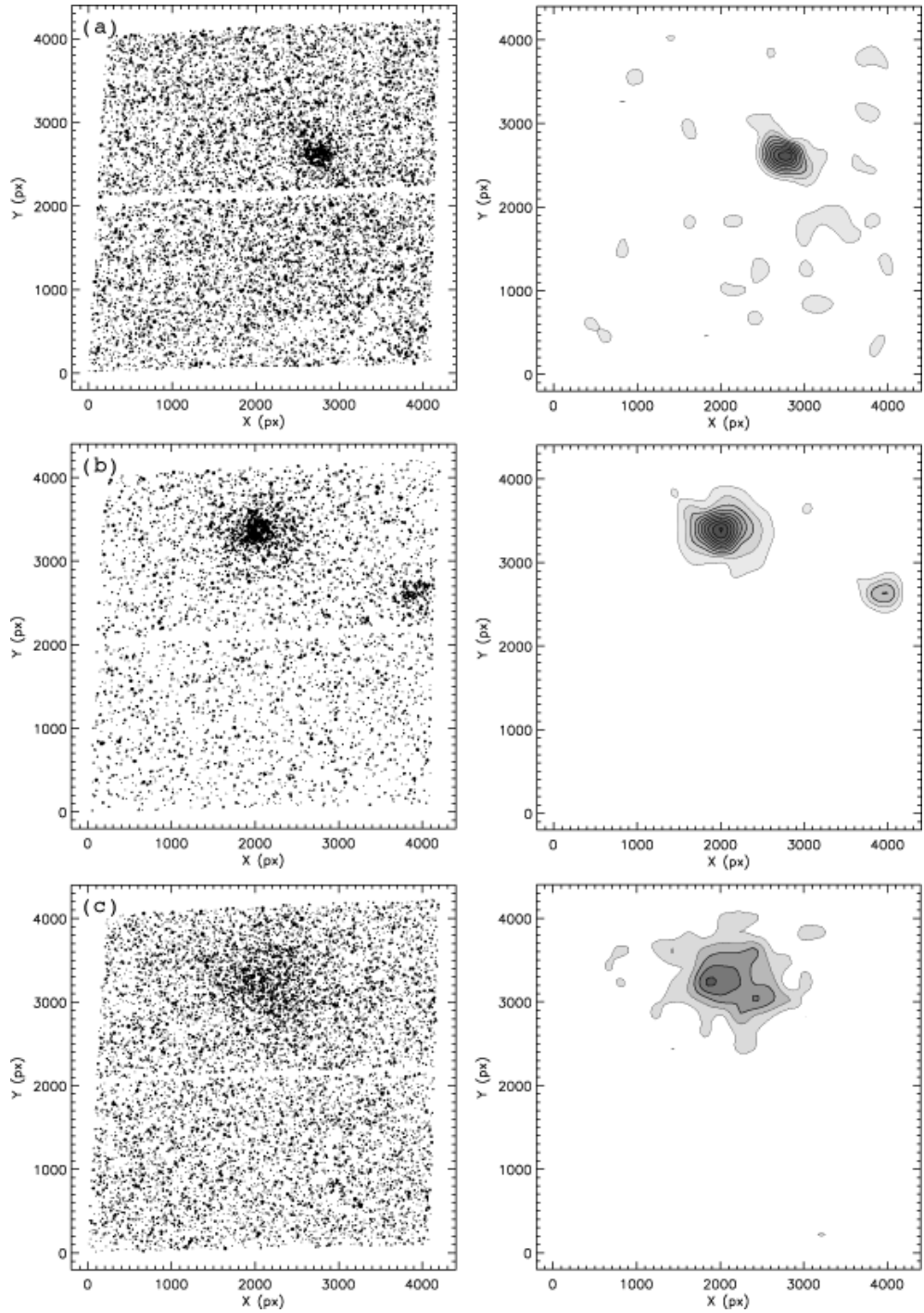}}
\caption{The observed ACS/WFC fields-of-view of the clusters (a) NGC~1983,
(b) NGC~2002 and (c) NGC~2010 (left panel), with the corresponding stellar 
density maps constructed with star counts on the photometric catalogs (right
panel). Symbol sizes correspond to the brightness of the marked stars. 
These maps demonstrate that the selected clusters show a variety in 
size, structure, stellar density and ambient local field (see text 
in \S~\ref{s:dynamics}).\label{f:maps}}
\end{figure*}

In order to assign a detection completeness fraction to each star,
we used {\sc dolphot} to perform a large number of artificial star tests. 
For each star in our final cleaned catalogue we first generated a list
of $150$ associated fake stars using the {\sc dolphot} utility 
{\sc acsfakelist}. We constrained these fake stars to lie close to
the real star on the (F555W, F814W) color-magnitude diagram (CMD) --
the fake stars were randomly and uniformly distributed in a box
of size $2\Delta_{V} \times 2\Delta_{C}$ centered on the real star,
where $\Delta_{V}$ was the maximum of $0.1$ mag and $3$ times the
uncertainty in the $m_{\rm 555}$ magnitude of the real star, and 
$\Delta_{C}$ was the maximum of $0.1$ mag and $3$ times the
uncertainty in the ($m_{\rm 555}-m_{\rm 814}$) color of the real star. 

We were also careful to constrain the spatial positions of the fake 
stars. Often this is done so that the fake stars lie at similar 
cluster-centric radii to the real star under consideration, but with 
unconstrained position angle. This saves on computation time because 
the calculations are then valid for groups of real stars at a time, 
rather than simply on an individual basis. However, because we are 
dealing with very young clusters in the present work, this implicit 
assumption of spherical symmetry is not necessarily valid. These objects 
could potentially still be embedded in gas or dust and thus exhibit a 
strongly varying background. They are also not very dynamically evolved, 
so may yet possess azimuthal asymmetries in stellar density as well as 
numerous randomly placed bright stars. Therefore, we adopted the more 
computationally expensive method of generating the list of fake stars 
locally around each real star on the CCD -- in this case randomly
and uniformly within a radius of $5\arcsec$. In doing this we were 
careful to account for the edges of the chips, including the 
inter-chip gap.

We then ran {\sc dolphot} in artificial stars mode on each list of
fake stars. When operating in this way {\sc dolphot} takes the next
fake star from the list, adds it to the input images, and then solves
for the position and photometric properties of the star using exactly
the same parameters as for the original (real star) analysis.
Once complete, we took the output photometry and ran it through our
quality filters, just as we did for the list of real stars.
Finally, we calculated the completeness fraction for a given real star 
by determining how many of its associated fake stars had been 
successfully `found'. We considered a fake star to be found if it
was recovered in both F555W and F814W and passed successfully through 
our quality filter, and provided its output position lay within $2$ 
pixels of its input position. Figure~\ref{f:compl} presents the 
completeness function of our photometry for all considered clusters in both
F555W (solid line) and F814W (dashed line) filters. 

\begin{figure}[t!]
\centerline{\includegraphics[clip=true,width=0.975\columnwidth]{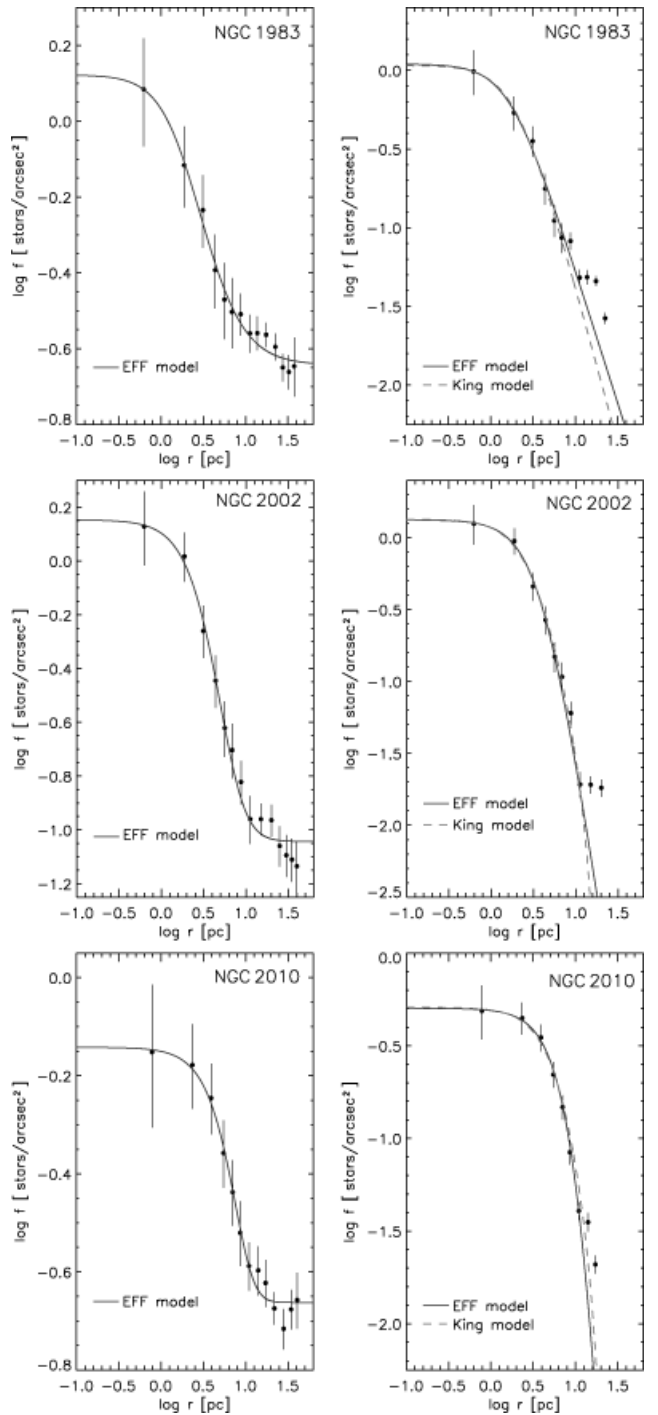}}
\caption{Radial surface density profiles of NGC~1983 (top) NGC~2002 (middle), and NGC~2010 
(bottom) with the best-fitting King and EFF models superimposed. The error bars represent 
Poisson statistics. {\em Left}: Density profiles of all detected stars, including 
the local field of the LMC. The best-fitting EFF model is over-plotted with a solid 
blue line. {\em Right}: Stellar surface density profiles after subtracting 
the background density level estimated from the best-fitting EFF model shown in the 
plots on the left. Both EFF and King models are overlaid. Since these clusters are not 
tidally truncated, the King models are used only tentatively for an estimation of a limiting 
radius for the clusters. Both models, however, seem to fit very well the density profiles at 
the inner parts of the clusters deriving comparable core radii. The derived 
structural parameters based on the best-fitting EFF (solid blue line) and King 
(dashed green line) models are given in Table~\ref{t:structpar}.
\label{f:prof}}
\end{figure}

\begin{deluxetable}{ccccc}
\label{t:param}
\tablecolumns{5}
\tablewidth{0pc}
\tablecaption{Characteristics of the clusters in our sample derived from
our photometry. \label{t:param}} 
\tablehead{
\colhead{Cluster} & 
\colhead{Size} &
\colhead{$f_{3\sigma}$} &
\colhead{Age} &
\colhead{$A_{V}$}\\
\colhead{} & 
\colhead{(pc)} &
\colhead{(stars~arcsec$^{-2}$)} &
\colhead{(Myr)} &
\colhead{(mag)} 
} 
\startdata
NGC~1983 &  ~7.3 & 0.72 & ~28& 0.17\\
NGC~2002 &  ~9.4 & 0.65 & ~18 & 0.19\\
NGC~2010 &  12.5 & 0.54 & 159 & 0.17\\
\enddata
\tablecomments{Sizes are derived from circular annuli encompassing the $3\sigma$ 
isopleth in the iso-density contour maps of the clusters in the observed regions. Stellar density 
$f_{3\sigma}$ is measured within these annuli and are given for comparison with the field
stellar density of Table~\ref{t:structpar}. Age and interstellar extinction for the clusters are 
estimated from evolutionary model fitting on the CMDs of stars within the $r_{\rm c}$ of the clusters. 
Sizes are given in physical  units after a conversion $1\arcmin \simeq 15$~pc assuming a distance modulus 
of $(m-M)_{0} \simeq 18.5$~mag \citep{clementini03,alves04,schaefer08}
for all clusters.}
\end{deluxetable}

\section{Cluster Morphologies and Structures}\label{s:dynamics}

\subsection{Morphology of the Clusters}\label{ss:contmap}

The stellar charts of the observed ACS/WFC fields of the clusters are 
shown in Figure~\ref{f:maps} (left panel). Our observations show clear 
differences in the appearance of the three clusters. 
In order to reveal the spatial distribution of the detected stars and the morphology 
of the clusters, we performed star counts on our 
photometric catalogs. The constructed `iso-density' contour maps for each region 
are also shown in Figure~\ref{f:maps} (right panel). The coordinate system used for these maps  
is that of the original ACS/WFC pixel coordinates. 
The star counts were performed in a quadrilateral grid divided in elements
with sizes $\sim$~190~$\times$~190 WFC~pix$^2$ each, which correspond to
about 10\arcsec~$\times$~10\arcsec\ (or $\approx$~2.5~pc~$\times$~2.5~pc at the distance of
the LMC).  The first isopleth in the contour map indicates the level of 1$\sigma$ 
above the mean background density, with $\sigma$ being the standard deviation 
of the background density. The subsequent levels are drawn in
steps of 1$\sigma$. All contours with density equal or higher than the
3$\sigma$ level, considered as the statistically significant density threshold, 
are drawn with thick lines. All three star clusters are revealed in these maps
as the dominant stellar concentrations in the observed regions. These maps
demonstrate that while NGC~1983 is a centrally concentrated compact cluster, NGC~2010 is 
a quite large and loose cluster, with NGC~2002 being an intermediate case of a large 
compact concentration. Moreover, NGC~2002 appears to be the most spherical cluster
in the sample, since NGC~1983 is more elliptical and NGC~2010 is more amorphous, 
for which, as indicated by the isopleths of density $\geq 3\sigma$, its most compact 
part is off-center. The sizes of the clusters as derived from circular annuli considered 
to encircle the $3\sigma$ isopleth for each cluster are given in Table~\ref{t:param}. 
The corresponding surface stellar density of each cluster is also given in this Table.

\subsection{Stellar Surface Density Profiles}\label{ssdp}

The observed fields cover quite large areas around the clusters in 
our sample, allowing the construction of the stellar surface density 
profiles of the clusters at relatively large distances from their 
centers. We divided the area of each cluster in concentric annuli of 
increasing steps outwards and we counted the number of stars
within each annulus. We corrected the counted stellar numbers for
incompleteness according to the corresponding completeness factors 
on a star-by-star basis. Completeness is a function of both distance from 
the center of the cluster and magnitude. Let $N_{i,c}$ be the 
completeness corrected stellar number within the $i$th annulus. We 
obtained the stellar surface density, $f_{i}$, by normalizing this number 
to the area of the corresponding annulus as $f_{i}=N_{i,c}/A_{i}$, with 
$A_{i}$ being the area of the annulus. The observed fields do not 
always fully cover the complete extent of the clusters, and, 
therefore, for the truncated annuli, we considered only the 
available area for the estimation of the corresponding surface density.

The ``raw'' radial stellar density profiles, meaning the 
surface stellar density as a function of distance from 
the center of each cluster, $f(r)$, show a smooth drop outwards 
away from the centers of the clusters. They drop to a uniform level, 
which represents the stellar density 
of the LMC field in the vicinity of the clusters, and it is measured 
by fitting the models of \cite{eff87} to the stellar surface density 
profiles, $\log{f(\log{r})}$, (Figure~\ref{f:prof}, left)  
as described below. In the case of NGC~2002 there is a small increase 
in the outskirts of the profile due to another neighboring small 
stellar concentration (see Figure~\ref{f:maps}). We do not consider 
this increase in our subsequent analysis on the structure of this
cluster. The errors in the profiles 
reflect the counting uncertainties, representing the Poisson statistics. 

\begin{deluxetable*}{ccccccc}
\label{t:structpar}
\tablecolumns{8}
\tablewidth{0.975\textwidth}
\tablecaption{Results of the application of both EFF and King
models fit to the observed radial surface density 
profiles of the clusters in our sample. \label{t:structpar}} 
\tablehead{
\colhead{} &
\multicolumn{4}{c}{\bf EFF Profiles} &
\multicolumn{2}{c}{\bf King Profiles}\\
\colhead{Cluster} & 
\colhead{$\alpha$} &
\colhead{$\gamma$} &
\colhead{$f_{\rm field}$} &
\colhead{$r_{\rm c}$} &
\colhead{$r_{\rm t}$} &
\colhead{$c$} \\
\colhead{} & 
\colhead{(pc)} &
\colhead{} &
\colhead{(stars~arcsec$^{-2}$)} &
\colhead{(pc)} &
\colhead{(pc)} &
\colhead{$\log{r_{\rm t}/r_{\rm c}}$} 
} 
\startdata
NGC~1983 &  1.68~$\pm$~0.28 & 1.69~$\pm$~0.32 & 0.23~$\pm$~0.02 & 1.89~$\pm$~0.43& 21.69 & 0.8\\
NGC~2002 &  3.92~$\pm$~0.68 & 3.91~$\pm$~1.00 & 0.09~$\pm$~0.01 & 2.56~$\pm$~0.61& 22.33 & 0.9\\
NGC~2010 & 13.32~$\pm$~6.03 & 9.81~$\pm$~8.04 & 0.21~$\pm$~0.01 & 5.19~$\pm$~3.34& 28.13 & 0.7\\
\enddata
\tablecomments{Parameters $\alpha$ and $\gamma$ are derived from the 
best-fitting EFF profile according to Eq.~(\ref{eq-elson}). Core radii, $r_{\rm c}$, 
are derived from the same models with Eq.~(\ref{eq-rc}). Tidal radii, $r_{\rm t}$, 
are measured with Eq.~(\ref{tidal}) from the best-fitting King models. Concentration
parameters, $c$, are estimated from the measured $r_{\rm t}$ and core radii 
derived from King models and Eq.~(\ref{core}). All radii are given in physical 
units after a conversion $1\arcmin \simeq 15$~pc assuming a distance modulus 
of $(m-M)_{0} \simeq 18.5$~mag \citep{clementini03,alves04,schaefer08}
for all clusters.}
\end{deluxetable*}

\subsection{Structural Parameters}\label{s:param}

We applied both the empirical model by \cite{king62} and the model of
\cite{eff87} (from here-on EFF) to the stellar surface density profiles 
of the clusters in order to obtain their structural parameters.
The EFF model is suited for clusters which are not tidally limited, while
King's empirical model represents tidally truncated clusters.  Both
models provide the opportunity to derive accurate characteristic radii
for the clusters. The core radius 
($r_{\rm c}$) describes the distance from the center of the cluster where the
stellar density drops to the half of its central value and the tidal radius 
($r_{\rm t}$) is the limit where the stellar density of the cluster drops to zero. 
The best-fitting EFF and King profiles are found by performing the 
Levenberg-Marquardt least-squares fit to the considered functions. This 
fit was performed with IDL\footnote{\url{http://www.ittvis.com/ProductServices/IDL.aspx}} 
(Interactive Data Language) with the use of the specially developed 
procedure MPFITFUN \citep{markwardt08}. For the
application of King's model the stellar surface density of the cluster
alone (with no contamination from the field) is necessary, while the EFF
model does not require any field subtraction. We first apply the latter
in order to estimate the core radius and the uniform background level
for each cluster. The subtraction of this density level from the measured 
surface density at each annulus gives the surface density profile of the 
cluster alone, from which we derive the core and tidal radii.

\subsubsection{Best Fitting EFF Profiles}\label{ss:effprof}

From studies of young LMC clusters by EFF it appears that these 
clusters are not tidally truncated. These authors developed a model more
suitable to describe the stellar surface density profile of such
clusters: \begin{equation} f(r)=f_{0}(1+r^{2}/{\alpha}^{2})^{-\gamma/2}+
f_{\rm{field}}, \label{eq-elson}\end{equation} where $f_{0}$ is the central stellar 
surface density, $a$ is a measure of the core radius and $\gamma$ is the 
power-law slope which describes the decrease of surface density of the cluster 
at large radii; $f(r)\propto r^{-\gamma/2}$ for $r\gg a$.  The uniform background density 
level is represented in Eq.~(\ref{eq-elson}) by $f_{\rm{field}}$. The measured 
parameters $\alpha$, $\gamma$ and $f_{\rm{field}}$, derived from the 
fitting procedure on our data are given for each cluster in Table~\ref{t:structpar}.
According to EFF model, the core radius, $r_{\rm c}$, is given from Eq.~(\ref{eq-elson}) 
assuming no contribution from the field as: \begin{equation} 
r_{\rm c}=\alpha(2^{2/\gamma}-1)^{1/2} \label{eq-rc}. \end{equation} The estimated 
core radii of the clusters derived from Eq.~(\ref{eq-rc}) are also given in Table~\ref{t:structpar}. 
The best-fitting EFF models are shown superimposed (blue lines) on the stellar surface
density profiles of Figure~\ref{f:prof}.

\subsubsection{Best fitting King Profiles}\label{ss:bestfitking}

In order to construct the density profiles of the clusters alone with no contribution 
from the field we subtract the background density level, $f_{\rm{field}}$, from the 
raw profiles of Figure~\ref{f:prof} (left). We then use the field-subtracted profile to 
derive an indicative tidal radius of the clusters as described by \cite{king62}.  
According to this model the density profile of a tidally 
truncated cluster is given as: \begin{equation} 
f(r)\propto\Bigl(\frac{1}{[1+(\frac{r}{r_{\rm c}})^2]^\frac{1}{2}}-\frac{1}
{[1+(\frac{r_{\rm t}}{r_{\rm c}})^2]^\frac{1}{2}}\Bigr)^{2}, \label{eq-king}
\end{equation} where $f$ is the stellar surface density, $r_{\rm c}$ and $r_{\rm t}$
the core and tidal radius, respectively and $r$ is the distance from the
center. The tidal radius, which represents the outskirts of the cluster, 
is found from the formula \begin{equation} f(r)=f_1(1/r-1/r_{\rm t})^2,\label{tidal} 
\end{equation} and the core radius, which describes the inner region of the cluster, 
is given from \begin{equation} f(r)=\frac{f_0}{1+(r/r_{\rm c})^2}. \label{core} \end{equation}
$f_0$ describes again the central surface density of the cluster and $f_1$ is a constant. 
The best-fitting King profiles apart from core and tidal radii for the clusters, deliver 
also the concentration parameters, $c$, defined as the logarithmic ratio of tidal to core radius 
$c=\log{(r_{\rm t}/r_{\rm c})}$, which refers to the compactness of the cluster. 
The derived $r_{\rm t}$ and $c$ for our clusters are given in Table~\ref{t:structpar}, 
while $r_{\rm c}$ derived from the best-fitting King profiles are omitted, as 
they are found in excellent agreement with those derived above 
from the EFF profiles.  The best-fitting King models of our clusters are shown 
with green dashed lines superimposed on the radial surface density profiles of 
Figure~\ref{f:prof} (right).

In general, both King and EFF models are in good agreement to each other in 
the inner and intermediate regions of the clusters, which are equally well fitted 
by both models. However, based on the fact that young compact clusters in the 
LMC are not tidally truncated, the EFF model should be considered as the best 
representative for them. On the other hand, the decontamination of the observed stellar 
samples from the populations of the local LMC field requires the determination 
of a limiting radius, $R_{\rm lim}$, for the clusters (\S~\ref{ss:fldsub}). Since 
EFF models do not predict such a radius, we treat the tidal radii derived from the 
application of King's theory as this radius for each cluster. It should be noted, though, that 
our clusters are not expected to be dynamically relaxed and therefore, the 
results on their tidal radii based on the application of King profile fitting 
should {\sl not} be taken literally but only tentatively for the determination 
of an indicative {\sl limit} radius for the clusters, and for reasons of comparison. 
For example, as seen in the charts of Figure~\ref{f:maps}, NGC~1983 cannot be directly 
distinguished from the rich stellar field in which it is embedded, as opposed, e.g., to NGC~2002, which
shows up as a compact concentration easily separated from its surroundings. 


\begin{figure*}[t!]
\centerline{\includegraphics[clip=true,width=0.875\textwidth]{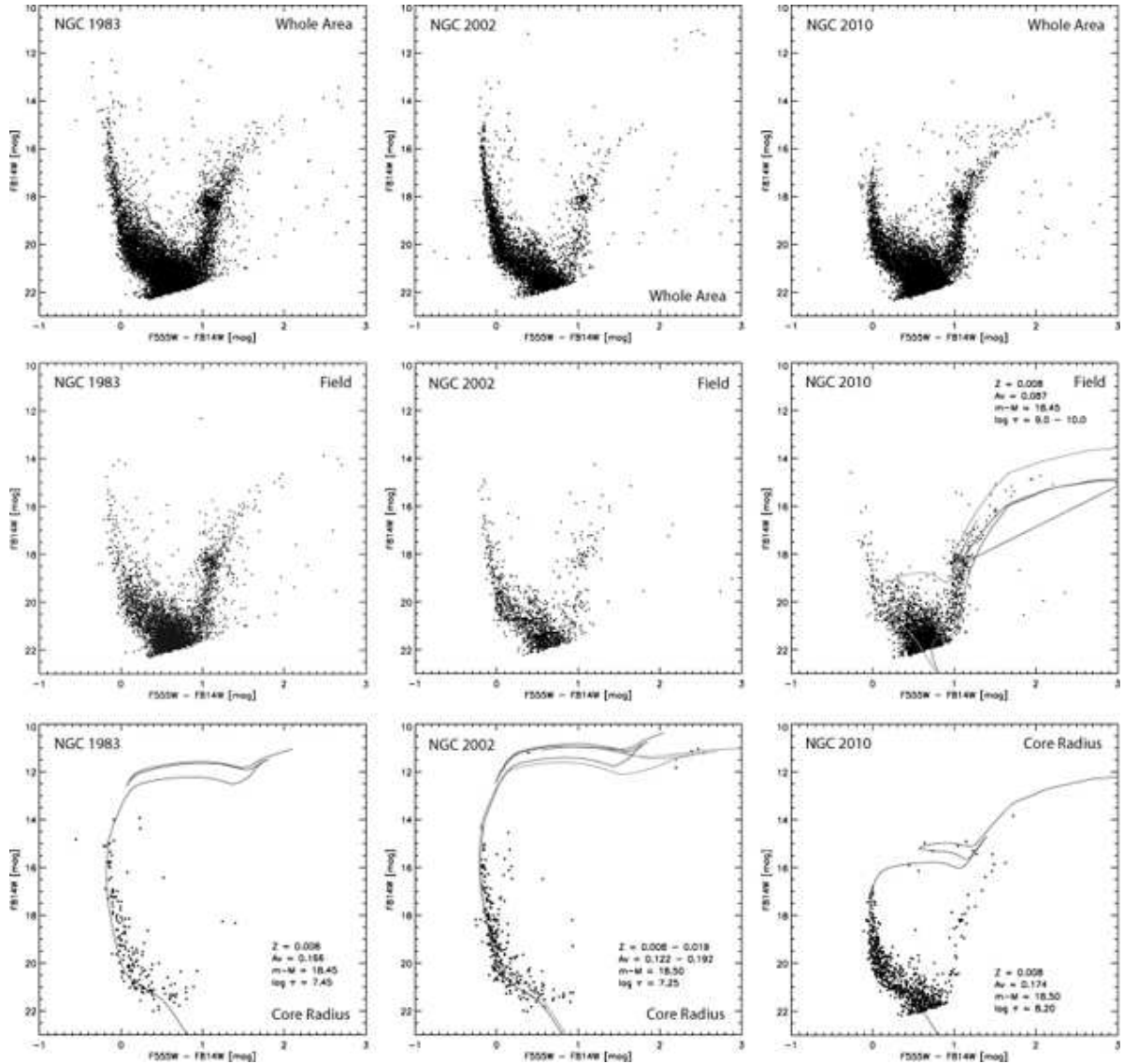}}
\caption{The m$_{\rm 555}-$~m$_{\rm 814}$ vs. m$_{\rm 814}$ CMDs 
of the stars detected with the best photometry (uncertainties $\sigma \leq$~0.1~mag in both 
filters) in the observed {\sl HST} ACS/WFC areas of the LMC clusters of our sample:
NGC~1983 (left), NGC~2002 (middle) and NGC~2010 (right). {\sl Top}: The complete CMDs
of all stars detected in the whole observed areas, which demonstrate the mixture 
of the young stellar populations of the clusters (MS) with those of their surrounding field, and 
the evolved ones of the general LMC field (RGB, RC). {\sl Middle}: The CMDs of the stars
located outside the  $r_{\rm t}$ of the clusters, representing the most probable populations 
of the local ambient of each cluster. {\sl Bottom}: The CMDs of the stars comprised within the $r_{\rm c}$ of
each cluster, which represent best the true CMDs of the  clusters. Isochrones from the
grid of evolutionary models by \cite{girardi02} are overlaid for an estimate of the 
age of the clusters and the old field (see \S~\ref{ss:cmds}).\label{f:totcmds}}
\end{figure*}

\section{Stellar Content of the Clusters}\label{s:starcont}

In the stellar charts of the three observed regions, shown in Figure~\ref{f:maps}
(left), the star clusters immediately appear as the most prominent stellar concentrations. 
These maps are constructed only from stars detected with photometric  uncertainties 
$\sigma \leq 0.1$~mag in both F555W and F814W filters. In our subsequent analysis 
we consider only these stars for each observed field. The ACS/WFC field-of-view of 
NGC~1983 covers 9~641 stars with good photometry, the one of NGC 2002 4~340, 
and that of NGC 2010 9~006 such stars.  

\subsection{Color-Magnitude Diagrams}\label{ss:cmds}

The $m_{555}-m_{814}$, $m_{814}$  Color-Magnitude Diagrams (CMDs)  of the total 
observed samples of stars for all three clusters are shown in Figure~\ref{f:totcmds} 
(top panel). These CMDs of the {\sl Whole Areas} demonstrate the existence of 
a significant mixture of different stellar populations in each region; from young stars, 
which cover the main sequence (MS) to the turn-off, belonging to  the clusters and 
their surrounding field, to the evolved Red Giant Branch (RGB) and Red Clump (RC)
stars, representing mostly the local background LMC field. The CMDs of Figure~\ref{f:totcmds} 
indicate that our photometry provides accurate magnitudes for stars 
with $m_{814}$~\lsim~22~mag. It is interesting to note that 
as seen from the CMDs of Figure~\ref{f:totcmds} (top), as well as from the stellar charts of Figure~\ref{f:maps}, 
the general field of the LMC in the observed regions shows to vary in stellar density. Specifically, 
both NGC~1983 and NGC~2010 are located in much denser stellar fields than NGC~2002, with 
the local field of NGC~1983 being particularly rich in red field stars. This is already demonstrated
by the measurements of the surface stellar density, $f_{\rm field}$, of the general field of all three 
clusters in  \S~\ref{s:param} (Table~\ref{t:structpar}; column 4).

The tidal radii, $r_{\rm t}$, of the clusters, as they are derived in \S~\ref{s:param}, basically 
define the spatial extent of each cluster, and therefore we treat  the  stars that are located at 
radial distances from the center of the clusters larger than $r_{\rm t}$ as the best representative 
stellar populations of the surrounding fields of the clusters. The corresponding CMDs, for 
which  we use the term {\sl Field CMDs}, are shown in  Figure~\ref{f:totcmds} (middle panel). 
In these CMDs the features of both RGB and RC of the general field of the LMC are well
apparent in all three observed regions. These populations are well documented as members 
of the LMC field in various HST studies of the star formation history (SFH) of this galaxy 
\citep[e.g.,][]{geha98, holtzman99, castro01, smeckerhane02}.  Specifically, the SFH of the 
LMC is found to be continuous with almost constant star formation rate (SFR) for over $\sim$~10~Gyr, 
with an increase of the SFR at around 1 - 4~Gyr ago \citep[for a summary see][their \S~3.1]{gouliermis06}.
This old LMC field population is present in all the Field CMDs of Figure~\ref{f:totcmds}. Since it covers 
the same parts in all three of them, we define the age-limits of these stars only on the CMD of 
the field of NGC~2010. With the use of the Padova grid of evolutionary models \citep{girardi02} 
for insignificant extinction and metallicity and distance typical for the LMC these limits are found 
to be $1~\lsim \tau/{\rm Gyr} \lsim~10$.

Moreover, the Field CMDs of Figure~\ref{f:totcmds} include also young MS stars, clearly suggesting that the general regions, 
where all three clusters are located,  include young stellar populations that  are not members of 
the clusters. This suggestion does not contradict previous studies on such clusters 
and their surroundings. Indeed, young compact clusters in the LMC are known to form together in larger structures of young stellar 
populations, the {\sl Stellar Aggregates} \citep[e.g.,][]{oey08}, the size of which seems to correlate 
with the duration of star formation in them \citep[e.g.,][]{efremov99}. Such clusters are not isolated 
as they may be formed in binary or multiple cluster systems \citep{dieball02}. An inspection
of the general regions where the clusters of our sample are located verifies that none of them
is an exception to this general rule.  Wide-field images of the regions around the clusters and the whole extent of the 
LMC in $B$, $R$ and $I$ were retrieved from the {\sl SuperCOSMOS Sky Survey}  \citep{hambly01}, 
available at the {\sl Aladin Sky Atlas} \citep{bonnarel00}. These images show that indeed all three 
clusters are members of larger young structures. Specifically, NGC~1983 is located at the central part of the 
star forming {\sl Shapley Constellation II}  \citep{shapley51, mckibbennail53} that coincides with the supergiant 
shell (SGS) {\sl LMC 3} \citep{meaburn80}, NGC~2002 belongs to the northern part of {\sl Constellation III} 
or SGS {\sl LMC 4}, and NGC~2010 is located in one of the eastern nebular filaments of SGS {\sl LMC 9}
in {\sl Constellation IX}.  

In order to compare these MS field populations with the actual stellar populations of the clusters 
we select the stars contained in the core radii of the clusters (see  \S~\ref{s:param}; Table~\ref{t:structpar}, 
column 5), as the most probable stellar members, and we construct the corresponding CMDs, which 
are shown in Figure~\ref{f:totcmds} (bottom panel).  A comparison of the Field CMDs with the CMDs of the 
clusters within their  core radii shows that there is a confusion between the young 
field populations and those of the clusters, especially for NGC~2002 and NGC~1983. The case of
NGC~2010 is more straightforward, as it can be seen that while the cluster is somewhat evolved (see 
discussion below), it seems that it is `embedded' in a younger field with brighter MS stars. On the other
hand, NGC~1983 and NGC~2002 are more complicated cases as both field and clusters seem to comprise 
the same type  of MS stars.  Taking into account the similarities of field stars to cluster members in the 
CMDs, the use of our method for the 
{\sl statistical} decontamination of the observed stellar samples from the field populations, discussed 
in \S~\ref{ss:fldsub}, becomes quite important. As we show in \S~\ref{ss:clusmem} this method allows 
us an accurate statistical determination of the most probable stellar members of each cluster.


\subsection{Ages of the Clusters}\label{ss:clusage}

We use the CMDs within the core radii of the clusters to derive an estimation of their ages. 
In the corresponding CMDs of Figure~\ref{f:totcmds} (bottom panel), the best-fitting isochrones from 
the Padova grid of evolutionary models \citep{girardi02} are also plotted. The process of 
isochrone fitting of the observed CMD results in 
an accurate measurement of the age of the system, providing that there are distinguished features in
the CMD to be fitted. Moreover, a good estimate of the distance and the interstellar reddening of the 
cluster is required. For clusters in the LMC, visual extinction is known to vary around very low values
\citep[e.g.,][]{degrijs02,gouliermis04,kerber06}, while the distance of the galaxy is also well defined
\citep[e.g.,][]{alves04,schaefer08}. Indeed, for our age estimation we had to vary little both of these parameters 
around typical values. Since the ACS photometric system is slightly different from the standard 
Johnson-Cousins system, we estimate the mean extinction in the F555W band, $A_{555}$, which we arbitrarily 
symbolize as $A_V$, using the computations by \cite{dario09}. These authors derive a conversion $R_{555} = 
A_{555}/E(m_{555}-m_{814}) \simeq 2.18$ and  $A_{555}/A_{814} \simeq 1.85$ assuming the extinction 
law of \cite{cardelli89} with $R_V\simeq 3.1$. In all cases we assume the typical metallicity of the LMC of $Z \simeq 
0.3$~-~$0.5$~$Z${\solar} \citep[e.g.,][]{westerlund97}, and therefore we use the models designed for $Z=0.008$. 
The derived ages and indicative visual extinction for all three clusters are given in Table~\ref{t:param}. Let us 
consider the results for each cluster individually:

{\sl (i) NGC~1983}: The CMD of the most prominent member stars of the cluster consists only of the MS, 
indicating that the cluster is quite young, but making the accurate determination of its age difficult. Indicatively, 
we apply the youngest isochrone that fits the brighter MS stars with $m_{814} \sim$~14~mag, which shows that the 
cluster has an age about 28~Myr. The selected isochrone fits best the blue part of the MS for a nominal
distance modulus $(m-M)_{0} \simeq 18.45$~mag and reddening of $E(m_{555}-m_{814}) \lsim 0.05$~mag, 
which corresponds to $A_V \simeq 0.17$~mag.

{\sl (ii) NGC~2002}: The CMD of NGC~2002 within the $r_{\rm c}$ of the cluster includes the 
significant number of five Red Supergiants (RSG) located at $2.2 \lsim (m_{555} - m_{814})$~\lsim~2.6~mag 
and 11~\lsim~$m_{814}$~\lsim~12~mag, which helps to constrain accurately the age of the system. 
However, the isochrone for metallicity $Z=0.008$ that seems to fit best this cluster with an age of 
$\tau \simeq$~18~Myr, cannot reach the RSG stars, which thus require the consideration of models 
of higher metallicity. Indeed the next available models from the Padova grid are those of almost solar metallicity 
of $Z=0.019$. The corresponding isochrone for the same age (plotted with a green line in Figure~\ref{f:totcmds}) does
fit the RSG, implying that either there is a significant change in the metallicity of the cluster, or that the evolutionary
models need to be fine-tuned for RSGs in low metallicities. In any case, our data give a fixed age for NGC~2002 of around
18~Myr. It should be noted that while the high metallicity model requires a modest $A_V \simeq 0.12$~mag to fit
the CMD, the low metallicity model predicts a higher extinction of $A_V \simeq 0.19$~mag. 

{\sl (iii) NGC~2010}: The CMD of Figure~\ref{f:totcmds} (bottom) for NGC~2010 clearly suggests that
this is the oldest cluster in our sample. The existence within $r_{\rm c}$ of bright stars of $m_{\rm 814} 
\simeq 15$~mag, evolved away from the MS allows us an accurate determination of the cluster age 
of $\tau \simeq 159$~Myr for $(m-M)_{0} \simeq 18.5$~mag and $A_V \simeq 0.17$~mag. However, 
the appearance in this spatially constrained CMD of a prominent RGB population, clearly suggests 
that the cluster may comprise a mixture of different  stellar populations.


\subsection{Field Decontamination}\label{ss:fldsub}

As we discuss earlier and seen from the CMDs of Figure~\ref{f:totcmds}, all three 
observed regions  suffer from significant contamination by field stars. With no information 
about cluster-membership probabilities for the observed stars, obtained, e.g., from radial 
velocities and/or proper motions, the quantitative decontamination of the clusters from the 
field stars on a statistical basis becomes a fundamental method to obtain the CMD of the
complete sample of true stellar members of each cluster. We decontaminate, thus, the 
observed CMDs from the contribution of the stellar members of the surrounding local field 
of the LMC with the use of a sophisticated random-subtraction technique.

This field-subtraction algorithm, originally designed by \cite{bonatto07}, is already 
developed in Paper~I, where it is also thoroughly described. According 
to this technique, the successful decontamination of the 
clusters from the field is based on the definition of the limiting radius, $R_{\rm lim}$, 
for each cluster. For the clusters in our sample, $R_{\rm lim}$ is given by the tidal 
radii, $r_{\rm t}$, estimated in \S~\ref{ss:bestfitking} with the use of King's theory. 
All stars located at distances $r>R_{\rm lim}$ from the center 
of the cluster are treated as field stars, while the rest (with $r \leq 
R_{\rm lim}$) are considered as the most probable cluster-member stars.  
Assuming a homogeneous field-star distribution, the number density of field 
stars is calculated in the Field CMD  (Figure~\ref{f:totcmds}; middle) constructed for 
stars with $r>R_{\rm lim}$, considering also the observational uncertainties 
in the photometry.

The decontamination process is then applied in three steps for each cluster: 
(i) Division of the CMDs of the stars with  $r \leq R_{\rm lim}$ (cluster region) 
and of those with $r > R_{\rm lim}$ (offset region), respectively, in 2D cells with 
the same axes along the m$_{\rm 814}$  and (m$_{\rm 555}-$m$_{\rm 814}$) 
directions. (ii) Calculation of the expected number density of field stars in each cell in the 
CMD of the offset region. (iii) Random subtraction of the expected number of
field stars from each cell from the CMD of the cluster region. 
The algorithm is applied several times using different cell sizes ($\Delta$m$_{\rm 814}$,
$\Delta$(m$_{\rm 555}-$m$_{\rm 814}$)) so that many different  decontamination 
results are derived. From these results we calculate the final probability that a star 
is identified as a `true' cluster member, and the final decontaminated CMD of the
cluster contains only the stars with the highest probability of being cluster members.  
This process allows us to minimize any artificial effects intrinsic to the method itself. 
A detailed description of the mathematical formulation of our method for field-subtraction 
is given in Paper I.  

\begin{figure*}[t!]
\centerline{\includegraphics[clip=true,width=0.975\textwidth]{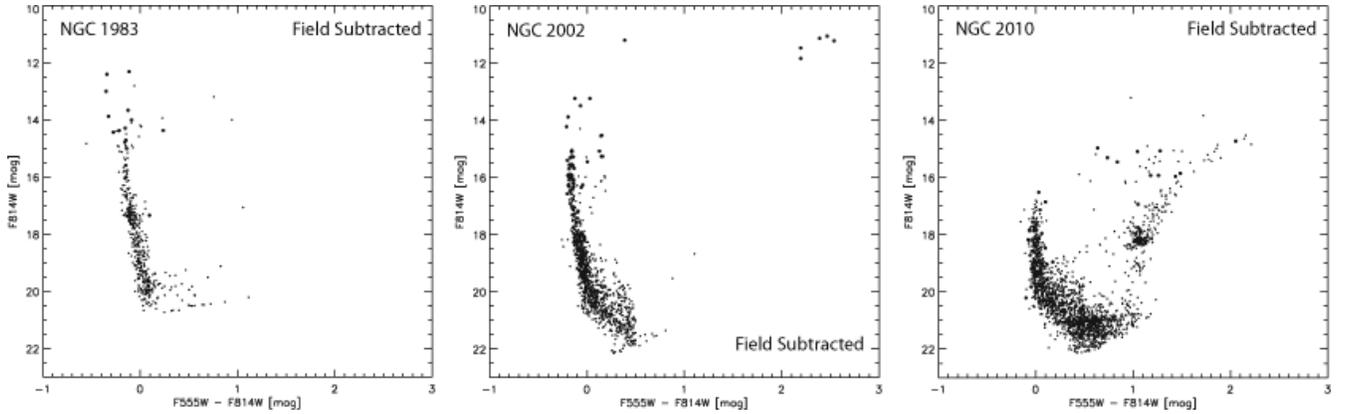}}
\caption{The m$_{\rm 555}-$~m$_{\rm 814}$ vs. m$_{\rm 814}$ CMDs of the stars retained  
as possible members of the clusters after the statistical removal of the contribution of the LMC field
from the originally observed CMDs of Figure~\ref{f:totcmds} (top panel). The Field CMDs of the 
offset regions around the clusters (Figure~\ref{f:totcmds}, middle panel) are used as representative
of the contaminating populations. Stars with probabilities $>$~70\% of being true members of the clusters 
(see \S~\ref{ss:fldsub}) are considered for these plots. Stars identified with 100\% membership are plotted 
with thick points.\label{f:syscmds}}
\end{figure*}

\subsection{Stellar Populations in the Clusters}\label{ss:clusmem}

We performed the field-subtraction process, as described above, 15 times for each 
cluster considering combinations for the CMD cell sizes of $\Delta$m$_{\rm 814}=$~0.5, 
1.0, 1.5, 2.5 and 3.0~mag and $\Delta($m$_{\rm 555}-$m$_{\rm 814})=$~1.0, 1.5 and 
2.0~mag. We obtained, thus, the probability for each star in each sample of being a 
member of cluster population. Since our method for diagnosis of stellar stratification is completeness-sensitive in the subsequent 
analysis we limit the sample of stellar members of all three clusters to those detected with our photometry with 
completeness $\geq$~70\%. The derived stellar samples comprise 424 stars
in NGC~1983, 1~102 stars in NGC~2002 and 2~265 stars in NGC~2010 with probability
larger than 70\% of being true cluster members. Figure~\ref{f:syscmds} shows the 
corresponding field-star-decontaminated CMDs of the cluster regions for all clusters. 
Stars found by the field-subtraction method with probability 100\% of being 
cluster members are plotted with thick points. From these CMDs it can be seen that 
our statistical subtraction of the CMD contamination by the field stars was efficient in
removing both young field populations of the immediate surroundings of the clusters,
{\sl and} old populations of the general LMC field.  Specifically, for NGC~1983 the MS 
stars remaining in the CMD of Figure~\ref{f:syscmds}, after the application of the field-subtraction, 
form a sharp narrow MS down to $m_{\rm 814} \simeq 21$~mag. This MS resembles that of the CMD within the $r_{\rm c}$ 
of the cluster shown in Figure~\ref{f:totcmds} (bottom), with no indication of the broad 
lower MS stars of the Field CMD of Figure~\ref{f:totcmds} (middle). The CMD of NGC~2002   
also demonstrates a very sharp MS similar to the $r_{\rm c}$ CMD of the cluster. The RSGs
located within the $r_{\rm c}$ of the cluster also remained in the CMD of Figure~\ref{f:syscmds}
with 100\% of membership probability. As far as NGC~2010 is concerned, 
the statistical field subtraction removed completely the bright MS young stars identified in 
\S~\ref{ss:cmds} as young field stars, while the older MS and evolved stars remained with high 
probabilities of being true cluster members.

It should be noted that the lower main sequence of the CMDs of the 
clusters may be also sensitive to any potential pre-main sequence (PMS) population that 
might exist in such young clusters. However, due to observational constrains, we would be 
able to detect only the turn-on of these PMS stars, which is known to coincide with the MSTO
\citep[see, e.g.,][]{nota06, gouliermis07}, very close to our detection limit. As a consequence,
any PMS population in the clusters would be seriously affected by the incompleteness in 
our photometry and therefore does not have any significance in the measured stellar numbers.
 Another bias that may affect the final numbers 
of low main sequence cluster-members is differential reddening that could displace stars
to the red-faint part of the CMD. However, differential reddening in young LMC clusters is
known to account for no more than $\Delta E(B-V) \simeq 0.1$ \citep[e.g.,][]{dirsch00, baume07},
which is well covered by the color uncertainties in our photometry at the low main sequence.
As a consequence, differential reddening is not considered in our treatment for field
subtraction.  It is interesting to note 
that while the Cluster CMDs of Figure~\ref{f:syscmds} for both
NGC~2002 and NGC~2010 show the same features with the corresponding $r_{\rm c}$ CMDs 
of the clusters (Figure~\ref{f:totcmds}, bottom), that of NGC~1983 shows clear indications of multiple
young populations with $m_{\rm 814} \lsim$~15~mag. Indeed, as we discuss in \S~\ref{ss:cmds}, 
multiplicity is a typical characteristic of young LMC clusters, and naturally it may lead to a merging 
process and mixing of different stellar populations within one cluster \citep[see e.g.,][]{portegieszwart07}. 
As we cannot verify or reject this scenario for NGC~1983, we treat all 424 stars that meet our membership 
criteria as true members of the cluster. Our diagnostic method for stellar stratification is applied 
in the following section on the final catalogs of stellar members of the clusters, with 70\% membership, 
the CMDs of which are shown in Figure~\ref{f:syscmds}.

\section{Diagnosis of Stellar Stratification in the Clusters}\label{s:methappl}

The {\sl effective radius} method for the diagnosis of stellar stratification is
based on the assumption that the observable counterpart of the {\sl Spitzer 
radius}, the {\sl effective radius}, of stars in a specific magnitude (or mass) 
range will be a unique function of this range if the system is segregated. 
In Paper~I the comparison of the results of this method with those from the 
`classical'  methods of the radial dependence of the clusters LFs and MFs, 
showed that the {\sl effective radius} method behaves more efficiently in the 
detection of the phenomenon,  providing direct proof of stellar stratification in 
star clusters without considering any model or theoretical prediction, and with 
no application of any functional fit. In this section we apply this method for the 
detection and quantification of stellar stratification to the young LMC clusters 
of our sample. 




\begin{figure}[t!]
\centerline{\includegraphics[clip=true,width=0.475\textwidth]{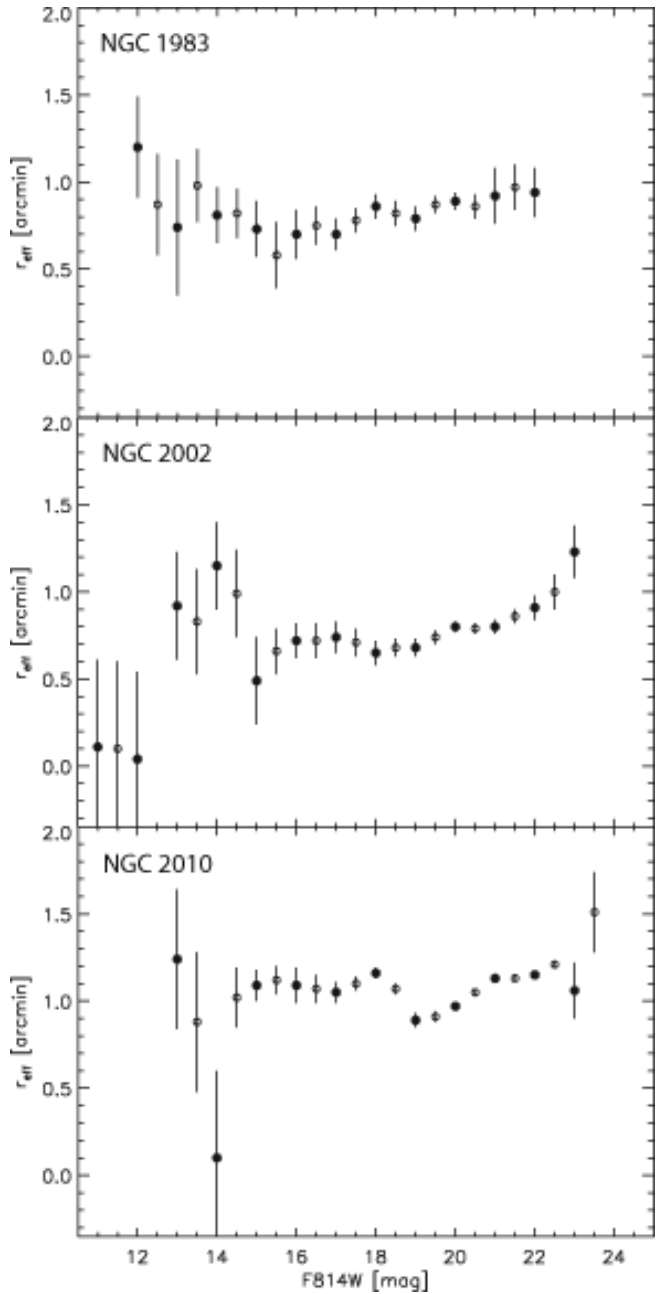}}
\caption{Estimated effective radii of stars within different magnitude ranges vs. the
corresponding mean magnitude for all three clusters.\label{f:reffmag}}
\end{figure}

\subsection{Effective Radii of the Star Clusters}\label{ss:calcreff}

We calculated the effective radii, $r_{\rm eff}$, of the stars in different magnitude
ranges, according to Eq.~(\ref{eq:reff}), using the completeness corrected numbers 
of stars for all three clusters. We performed this calculation only for the star-members of the 
clusters, as they are established in \S~\ref{ss:clusmem}  (CMDs of Figure~\ref{f:syscmds}), 
and we used their brightness measured in the F814W filter, since this filter 
provided the most complete and accurate photometry (see \S~\ref{ss:phot}). In order to 
identify any functional relation between $r_{\rm eff}$ and the corresponding 
brightness range, we binned the stars according to their magnitudes. Since 
the measurement of $r_{\rm eff}$ depends strongly to the number of stars per bin, 
this process naturally inherits systematic uncertainties to the derived relations, 
because of the fact that some stars could belong to neighboring bins due to 
photometric uncertainties. Therefore, we performed the same binning process per cluster 
several times by changing the bin sizes and/or the magnitude limits, within which the
stars are binned. We found that a reasonable bin size, which provides good statistics 
and allows for the detailed observation of the dependence of $r_{\rm eff}$ on magnitude, 
is that of 0.5~\lsim~$\Delta m_{814}$~\lsim~1.0~mag. For this range of bin sizes we did 
not find any significant change in the general trend of  $r_{\rm eff}(m_{\rm 814})$, which is 
also found to be independent of the selected magnitude limits for binning the stars. 

In Figure~\ref{f:reffmag} we indicatively show the relations between the computed 
effective radii and the corresponding magnitude bins derived from two different binning 
processes with bin sizes of 1~mag and magnitude ranges shifted by 0.5~mag from 
each other. The combination of both distributions with these specific binning 
characteristics (each plotted with different symbols) provides measurements of 
$r_{\rm eff}$  for every 0.5~mag range. The derived trends of $r_{\rm eff}(m_{\rm 814})$, 
shown in these plots, are not different from the corresponding distributions derived 
for $\Delta m_{814}=0.5$~mag,  except of being somewhat smoother. In general the 
relations of  Figure~\ref{f:reffmag} are representative of those found from any applied 
binning, and therefore we consider them in the following section for the derivation 
of our results concerning stellar stratification in the clusters. 

\subsection{Stellar Stratification in the Star Clusters}\label{ss:methres}

The $r_{\rm eff}(m_{\rm 814})$ graphs of Figure~\ref{f:reffmag} are plotted 
on the same scale for reasons of comparison between the clusters. In these 
plots it is shown that the effective radius does behave as a function of 
magnitude, but  not necessarily providing proof of stellar stratification
for all clusters. A definite indication of stellar stratification would require
the effective radii of the faintest stars to be systematically larger than those
of the brightest ones for the whole observed brightness range. This is not the 
behavior we observe at least for two of the three clusters of our sample. 
Indeed, as we discuss in Paper~I, $r_{\rm eff}$ is not a monotonic function
of brightness, and mass segregation, {\sl if any}, is not expected to behave 
in the same manner in every cluster, but rather to depend on the intrinsic 
stellar characteristics of each of them. For example, while in NGC~2002 
there is a trend of fainter stars having systematically larger effective radii 
for the whole luminosity range, this is not the case for NGC~1983 that shows 
this trend for only a limited magnitude range, and certainly not for 
NGC~2010, which one may argue is not even segregated.

Specifically, for NGC~1983, the brightest magnitude bins, down to $m_{\rm 814} 
\simeq$~16~mag, seem to include stars that are located away from the central 
part of the cluster. Indeed, the brightest stars in the Cluster CMD 
(Figure~\ref{f:syscmds}) with $m_{\rm 814}  \lsim$~15.5~mag are found to 
be distributed within the largest  $r_{\rm eff}$ of about 1\farcm2 away 
from the center of the cluster and not closer than about  0\farcm6. On the other hand, 
taking into account the remaining magnitude bins of $m_{\rm 814} \geq$~15.5~mag, 
the graph of $r_{\rm eff}(m_{\rm 814})$  for NGC~1983 provides clear indications that 
the cluster is segregated with an almost monotonic dependency of larger
$r_{\rm eff}$ to fainter stars. 

The graph for NGC~2002 is quite different, giving evidence that the cluster is 
segregated in the whole extend of observed stellar luminosities. There is, however, 
a dependency of the appearance of a relation between $r_{\rm eff}$ and brightness
on the considered magnitude range. Indeed, while the three brightest magnitude bins, 
comprising six stars (the five RSG and the brightest MS star), correspond to the smallest 
effective radii, the next fainter bins, which include six more MS stars,  show that these stars 
are located well away from the center of the cluster. Moreover, from the plot of 
Figure~\ref{f:reffmag} for the fainter stars, it is almost self-evident that they are segregated 
as the corresponding relation $r_{\rm eff}(m_{\rm 814})$ demonstrates an increase of 
$r_{\rm eff}$ with brightness.

The behavior of this relation for NGC~2010 is quite different from both 
the previous cases, as there is no general trend for larger effective radii with 
fainter magnitudes observable, and thus the cluster behaves as if there is no 
significant stratification occurring. Specifically, apart from the fact that the three brightest
magnitude bins are inconclusive, for more faint stars there is a ``plateau'' in 
the function  $r_{\rm eff}(m_{\rm 814})$,
which appears rather flat for magnitudes down to $m_{\rm 814} \simeq$~19~mag,
while for even fainter stars, down to the faintest observed magnitude, the dependence
of  $r_{\rm eff}$ over $m_{\rm 814}$ appears again, giving evidence of mild mass 
segregation for the faintest stars at the outskirts of the cluster.

\begin{deluxetable*}{cccccc}
\label{t:segdeg}
\tablecolumns{5}
\tablewidth{0pc}
\tablecaption{Magnitude limits, the corresponding effective radii
limits, and the derived slopes $S_{\rm strat}$ indicative of the degree of
stratification for selected brightness ranges for all three clusters of our sample.
\label{t:segdeg}} 
\tablehead{
\colhead{Cluster} & 
\colhead{$m_{\rm 814}$} &
\colhead{$\Delta m_{\rm 814}$} &
\colhead{$r_{\rm eff}$} &
\colhead{$\Delta r_{\rm eff}$}&
\colhead{$S_{\rm strat}$}\\
\colhead{} & 
\colhead{(mag)} &
\colhead{(mag)} &
\colhead{(arcmin)} &
\colhead{(arcmin)} &
\colhead{(mag/arcmin)} 
} 
\startdata
NGC~1983  & 12.0~-~15.5 &  3.5  & 1.20~-~0.58 &  $-0.62$ &  $-0.177$\\
                       & 15.5~-~22.0 &  6.5  & 0.58~-~0.94 &  $+0.36$ &  $+0.055$\\
                       & 12.0~-~22.0 & 10.0 & 1.20~-~0.94 &  $-0.26$ &  $-0.026$\\
NGC~2002 & 11.0~-~14.5 & 3.5 & 0.11~-~0.99 &  $+0.88$  &  $+0.251$\\
                       & 15.0~-~23.0 & 8.0 & 0.49~-~1.23 & $+0.74$  &  $+0.093$\\
                       & 11.0~-~23.0 & 12.0 & 0.11~-~1.23 & $+1.12$  &  $+0.093$\\
NGC~2010  & 13.0~-~14.0 & 1.0 & 1.24~-~0.10 & $-1.14$ &  $-1.140$\\
                       & 14.5~-~18.5 & 4.0 & 1.02~-~1.07 & $+0.05$  &  $+0.013$\\
                       & 19.0~-~23.5 & 4.5 & 0.89~-~1.51 & $+0.62$  &  $+0.138$\\
                       & 13.0~-~23.5 & 10.5  & 1.24~-~1.51 & $+0.27$&  $+0.026$\\
\enddata
\tablecomments{Negative values of $S_{\rm strat}$ or values close to zero indicate non-segregated
clusters at least within the specific magnitude limits (see \S~\ref{ss:methres}). The third row for
every cluster refers to the whole observed magnitude range.}
\end{deluxetable*}

\subsection{Degree of Stellar Stratification}\label{ss:stratdeg}

In order to parametrize the phenomenon of stellar stratification as it exhibits (or not) 
itself in the considered clusters, we assume as null hypothesis that all clusters 
are segregated, but to a different  {\sl degree}. This is evident, as discussed above, 
from the $r_{\rm eff}$ vs. magnitude plots of Figure~\ref{f:reffmag}. This {\sl  degree 
of stratification} in each cluster may be, thus,  expressed in terms of the brightness 
range of its stars, and the effective radius within which they are confined. As a 
consequence, in order to quantify stellar stratification as observed in every cluster, 
one may use the two primary output parameters of the effective radius method, 
i.e., the {\sl magnitude range} and the {\sl corresponding effective radii} of 
segregation. These parameters actually represent the ranges in magnitudes 
and radial distances that define {\sl the slope} of the observed relation 
$r_{\rm eff}(m_{\rm 814})$. Naturally, for a non-segregated cluster this slope,
$S_{\rm strat}$, should be $\leq 0$, as it is derived from the graphs of Figure~\ref{f:reffmag}. 
Considering that a steeper positive slope of this relation represents {\sl a higher 
degree of stratification}, we parametrize stellar stratification in 
the clusters  by measuring $S_{\rm strat} \equiv \Delta r_{\rm eff}$/$\Delta m_{\rm 814}$
and quantify the phenomenon through this {\sl  degree}.

As mentioned above, $r_{\rm eff}(m_{\rm 814})$ does not represent a systematic 
relation throughout the whole observed magnitude range in every cluster, but, as seen 
in Figure~\ref{f:reffmag}, it shows differences depending on the selected magnitudes. 
For every cluster we select one indicative magnitude limit to separate the bright from 
the faint stars and measure the corresponding degree of stratification. This selection 
is based on the magnitude bin, where we observe dramatic changes in the relation 
$r_{\rm eff}(m_{\rm 814})$ in each cluster. As a consequence, we select two magnitude 
ranges, one spanning from the brightest magnitude bin down to the selected
limit, and the other starting at this limit until the faintest magnitude bin, 
and we estimate their corresponding $S_{\rm strat}$ for all three clusters. These magnitude 
ranges, the corresponding $\Delta m_{\rm 814}$, and $\Delta r_{\rm eff}$, and the derived 
$S_{\rm strat}$ are given in Table~\ref{t:segdeg}, along with the derived $S_{\rm strat}$
for the whole observed magnitude range for every cluster. The values of $S_{\rm strat}$,
given in this table, correspond to the {\sl degree of stratification} for each cluster in every 
given brightness range. 

\subsection{Results and Discussion}\label{ss:results}

From the $S_{\rm strat}$ measurements of Table~\ref{t:segdeg} one can see that 
while indeed the bright stars of NGC~1983 are {\sl not
segregated}, stars with $m_{\rm 814}$~\gsim~15.5~mag show a definite trend in their
relation $r_{\rm eff}(m_{\rm 814})$, indicative of stratification with a degree of $+0.06$. On 
average, however, the cluster does not appear to demonstrate stellar stratification. We 
characterize, thus, NGC~1983 as {\sl partially} segregated cluster. On the other hand, NGC~2002 
shows a definite trend of larger $r_{\rm eff}$ for fainter stars within both selected magnitude ranges,
as well as for the whole extend of observed magnitudes. It is a definite case of stellar stratification 
with a degree of stratification of $+0.09$. Finally, NGC~2010 is the most peculiar case, as the results
for the few brightest stars are quite inconclusive, and for the fainter stars with 14.5~\lsim~$m_{\rm 814}$~\lsim~18.5~mag
the cluster appears slightly segregated, with a relation  $r_{\rm eff}(m_{\rm 814})$ that is rather flat. 
On the other hand, the most prominent trend in this relation appears for even fainter stars with 
$m_{\rm 814}$~\gsim~18.5~mag and a degree of stratification $S_{\rm strat} \simeq +0.14$. 
However, if indeed the cluster was segregated for this magnitude range, the observed trend 
should appear as a continuation of the (non stratified) relation $r_{\rm eff}(m_{\rm 814})$ of 
the brighter stars. More specifically the $r_{\rm eff}$, where the trend starts for $m_{\rm 814} 
\simeq$~19~mag ($\sim$~0\farcm9) is {\sl smaller} than that of the faintest non segregated stars 
of $m_{\rm 814} \simeq$~18.5~mag with $r_{\rm eff} \sim 1\farcm1$. This behavior of the 
relation $r_{\rm eff}(m_{\rm 814})$  corresponds to fluctuations in the spatial distribution
of the faint stars around the center of the cluster rather than to true segregation. As a consequence,
we characterize NGC~2010 as {\sl not segregated} cluster.

While the aforementioned results on three clusters cannot be used for a statistically complete 
interpretation of the phenomenon of stellar stratification in young LMC clusters in connection 
to their individual characteristics, it is worthwhile to search for any dependence of the degree of
stratification to the structural and evolutionary parameters we derived for each cluster in our sample.
Concerning the structural parameters shown in Table~\ref{t:structpar}, all clusters
are found with comparable concentration parameters, and tidal radii, $r_{\rm t}$, that are not far from each 
other. The most important difference we find in these parameters is the core radius, $r_{\rm c}$, of NGC~2010, 
the non (or less) segregated cluster, which is at least two times larger than those of the other two 
clusters. In addition, from the characteristics of Table~\ref{t:param} it is shown that this cluster is
the most extended and least dense in the sample. Moreover, the CMD of the cluster 
(Figure~\ref{f:syscmds}) shows indications of multi-age stellar populations, suggesting that 
NGC~2010 is a cluster with multiple stellar generations. In such clusters a large fraction of the first generation
of stars is lost early in the cluster evolution due to its expansion and stripping of its outer layers resulting
from early mass loss associated \citep[e.g.,][]{dercole08}. In general, the loose appearance of a 
low-density cluster with large $r_{\rm c}$, comes to the support of this speculation. While this 
cannot be verified with the present data, it would be reasonable to suggest that the lack of stellar 
stratification in this cluster will make the effect of early mass loss due to stellar evolution less 
destructive leading to longer lifetimes \citep{vesperini08}.
 
As far as NGC~2002 is concerned, it is interesting to note that this cluster, being the most segregated in 
the sample, is actually the youngest (Table~\ref{t:param}) and located in a region, which appears to 
correspond to the emptiest general LMC field of the three (Table~\ref{t:structpar}). Naturally, these 
observations allow us to suggest that detection of stellar stratification may have been benefitted by the
loose contaminating stellar field, and that the observed stellar stratification in NGC~2002 may be 
{\sl primordial in nature} due to the star formation process, or due to very early dynamical 
evolution. Indeed recent studies show that  cool, fractal clusters can dynamically mass segregate 
on timescales comparable to their crossing-times, far shorter than usually expected 
\citep[e.g.,][]{allison09b}. The exact nature, though, of the observed segregation cannot be fully 
understood without detailed kinematic information about the cluster members, or deeper photometry
that will provide the  complete IMF of the cluster. Finally, the observed {\sl partial} stellar stratification 
in NGC~1983 is driven by the fact that the brightest stars of the cluster are distributed in a wide range 
of distances from the cluster center. In general, the behavior of $r_{\rm eff}$ as function of stellar 
magnitude cannot be connected to any specific difference of its characteristics from those 
of the other clusters, apart from the fact that the cluster belongs to the richest region in field populations, 
and it is a rather elliptical cluster.

\section{Conclusions}\label{s:concl}

We present a coherent comparative investigation of stellar stratification in
three young LMC clusters with the application of a robust method for the assessment
of this phenomenon on deep HST imaging. In a cluster where 
stellar stratification occurs, the segregated brighter
stars are expected to be more centrally concentrated than the non-segregated
fainter stars. Stellar stratification can, thus, be investigated from the
dependence of the radial extent of stars in specific magnitude (mass) ranges on
the corresponding mean magnitude (mass). According to this notion, in Paper~I
\citep[][]{gouliermis09} we developed and verified the efficiency of a robust
method for the assessment of stellar stratification in star clusters. This method
is based on the calculation of the mean-square radius, the {\sl effective radius}
$r_{\rm eff}$, of stars in different magnitude ranges, and the investigation of its
dependence on magnitude as indication of stellar stratification in the cluster.

In the present study we apply the {\sl effective radius method} for the detection
and quantification 
of stellar stratification in young star clusters in the LMC. We select three
clusters observed with the high resolving efficiency of HST/ACS, specifically
NGC~1983, NGC~2002 and NGC~2010, on the basis of the differences in their
appearance, structure, stellar content, and surrounding stellar field. Our
photometry delivered complete stellar catalogs down to $m_{\rm 814} \simeq 22$~-~23~mag
for all three observed fields (\S~\ref{ss:phot}). The stellar surface density maps 
of the observed regions, constructed from star counts on the photometric catalogs, 
demonstrate that all three clusters are morphologically quite different 
from each other. NGC~1983 appears to be
a centrally concentrated, rather elliptical compact cluster, NGC~2002 a 
compact, spherical stellar concentration, and NGC~2010 a large, loose and rather
amorphous cluster (\S~\ref{ss:contmap}).

The application of both EFF \citep{eff87} and King's \citep{king62} models on the
stellar surface density profiles of the clusters allowed us the
measurement of the background stellar field density of the regions, and the
estimation of the core radii, $r_{\rm c}$, of the clusters. While these clusters
do not seem to be tidally truncated, a tidal radius, $r_{\rm t}$ is estimated,
indicative of the limiting radius of each cluster (\S~\ref{s:dynamics}). The
latter is essential for the successful application of a random field subtraction
technique for the decontamination of the stellar samples of the clusters from
their surrounding young stellar ambient and the general LMC field
(\S~\ref{ss:fldsub}). We derive the ages of the clusters from the
stellar populations comprised within the $r_{\rm c}$ of the clusters and we find
ages of $\tau \simeq$~28~Myr, 18~Myr and 159~Myr for NGC~1983, NGC~2002, and
NGC~2010 respectively (\S~\ref{ss:clusage}).

We finally apply the {\sl effective radius method} for assessing stellar stratification
in the clusters and for its quantitative study (\S~\ref{s:methappl}). We bin the stars according to their 
magnitudes in the F814W filter, and we estimate the corresponding effective radii, 
$r_{\rm eff}$, for every cluster in our sample. We then plot the derived radii versus 
the corresponding mean magnitude and we investigate the functional relation, 
$r_{\rm eff}(m_{\rm 814})$, between these parameters. With our method it is shown that 
stellar stratification behaves differently in every cluster. NGC~1983 appears to 
be {\sl partially segregated}, since its brightest stellar content does not appear 
to be centrally concentrated, while its fainter stars show a mild dependency of 
$r_{\rm eff}$ towards lower values for fainter magnitudes. The results on the 
bright stars of NGC~2010 show evidence of lack of stratification, since
the brightest stars are located quite far away from its center. As far as the 
faint stars are concerned they show a rather flat $r_{\rm eff}(m_{\rm 814})$
relation, with no evidence of segregation in this relation for the faintest stars. 
Consequently, this cluster is characterized as {\sl not segregated}.
Finally, NGC~2002 is found to be well segregated for both its brighter and fainter 
stars. It is a clear case of proof of stellar stratification with the effective radius method.

We propose the {\sl slope} $S_{\rm strat} \equiv \Delta r_{\rm eff}$/$\Delta m_{\rm 814}$, 
measured for stars in selected magnitude ranges, as well as in the whole magnitude range, as the most appropriate 
parameter for the quantification of the {\sl degree of stratification}, and we use it to characterize the 
phenomenon as it is observed in the clusters. From the derived values of $S_{\rm strat}$ 
(\S~\ref{ss:stratdeg}) we find that NGC~2002 is the most strongly segregated cluster in the sample
with a degree of stratification $S_{\rm strat} \simeq +0.09$, NGC~1983 is a partially segregated 
cluster with $S_{\rm strat} \simeq +0.06$ for the faintest stars with $m_{\rm 814}$~\gsim~15.5~mag, 
and NGC~2010 is not segregated with indicative $S_{\rm strat} \simeq +0.03$ (\S~\ref{ss:results}). 
Finally, it should be noted that the extension 
of this study to a larger sample of LMC clusters observed with HST will provide a more 
complete picture of the phenomenon of stellar stratification. Naturally, the development 
of such a comparative scheme, which will include star clusters in the whole extent of the 
evolutionary sequence, requires a comprehensive set of ACS and, in the near future, 
WFC3 observations.

\acknowledgements

D. A. G. kindly acknowledges the support of the German Research
Foundation (Deu\-tsche For\-schungs\-ge\-mein\-schaft, DFG) through 
grant GO~1659/1-2, and the German Aerospace Center (Deutsche Zentrum 
f\"{u}r Luft- und Raumfahrt, DLR) through grant 50~OR~0908. Sincere 
acknowledgements go also to M. Kontizas and E. Kontizas the collaboration with 
whom originally planted the seed of the idea for the Effective Radius method.
Based on observations made with the NASA/ESA {\em Hubble Space Telescope}, obtained from the
data archive at the Space Telescope Science Institute. STScI is operated
by the Association of Universities for Research in Astronomy, Inc. under
NASA contract NAS 5-26555. 

Facilities: HST


\begin{thebibliography}{}

\bibitem[Allison et al.(2009a)]{allison09} 
Allison, R.~J., Goodwin, S.~P., Parker, R.~J., 
Portegies Zwart, S.~F., de Grijs, R., \& Kouwenhoven, 
M.~B.~N.\ 2009a, \mnras, 395, 1449

\bibitem[Allison et al.(2009b)]{allison09b} 
Allison, R.~J., Goodwin, S.~P., Parker, R.~J., de Grijs, R., Portegies Zwart, S.~F., 
\& Kouwenhoven, M.~B.~N.\ 2009b, \apjl, 700, L99 

\bibitem[Alves(2004)]{alves04} 
Alves, D.~R.\ 2004, New Astronomy Review, 48, 659 

\bibitem[Ascenso et al.(2009)]{ascenso09} 
Ascenso, J., Alves, J., \& Lago, M.~T.~V.~T.\ 2009, \aap, 495, 147 

\bibitem[Baume et al.(2007)]{baume07} 
Baume, G., Carraro, G., Costa, E., M{\'e}ndez, R.~A., \& Girardi, L.\ 2007, \mnras, 375, 1077 

\bibitem[Bica et al.(1999)]{bica99} 
Bica, E.~L.~D., Schmitt, H.~R., Dutra, C.~M., \& 
Oliveira, H.~L.\ 1999, \aj, 117, 238 

\bibitem[Bica \& Schmitt(1995)]{bica95} 
Bica, E.~L.~D., \& Schmitt, H.~R.\ 1995, \apjs, 101, 41 

\bibitem[Bonatto \& Bica(2007)]{bonatto07}
Bonatto, C., \& Bica, E.\ 2007, \mnras, 377, 1301 

\bibitem[Bonnarel et al.(2000)]{bonnarel00} 
Bonnarel, F., et al.\ 2000, \aaps, 143, 33 

\bibitem[Bonnell \& Davies(1998)]{bonnell98} 
Bonnell, I.~A., \& Davies, M.~B.\ 1998, \mnras, 295, 691 

\bibitem[Bonnell \& Bate(2006)]{bonnell06} 
Bonnell, I.~A., \& Bate, M.~R.\ 2006, \mnras, 370, 488 

\bibitem[Brandl et al.(1996)]{brandl96} 
Brandl, B., et al. 1996, \apj, 466, 254

\bibitem[Cardelli et al.(1989)]{cardelli89} 
Cardelli, J.~A., Clayton, G.~C., \& Mathis, J.~S.\ 1989, \apj, 345, 245 

\bibitem[Castro et al.(2001)]{castro01} 
Castro, R., Santiago,  B.~X., Gilmore, G.~F., Beaulieu, S., \& Johnson, R.~A.\ 2001, \mnras, 326, 333 

\bibitem[Chandrasekhar (1942)]{chandrasekhar42} 
Chandrasekhar, S.\ 1942, Principles of Stellar 
Dynamics. Reprint 2005 by Dover Pubs. (New York). 
ISBN-13 9780486442730

\bibitem[Clementini et al.(2003)]{clementini03} 
Clementini, G., Gratton, R., Bragaglia, A., Carretta, E., Di Fabrizio, L., 
\& Maio, M.\ 2003, \aj, 125, 1309 

\bibitem[D'Ercole et al.(2008)]{dercole08} D'Ercole, A., 
Vesperini, E., D'Antona, F., McMillan, S.~L.~W., 
\& Recchi, S.\ 2008, \mnras, 391, 825 

\bibitem[Da Rio, Gouliermis \& Henning(2009)]{dario09} 
Da Rio, N., Gouliermis, D.~A., \& Henning, T.\ 2009, \apj, 696, 528 

\bibitem[de Grijs et al.(2002)]{degrijs02} de Grijs, R., Gilmore, 
G.~F., Johnson, R.~A., \& Mackey, A.~D.\ 2002, \mnras, 331, 245 

\bibitem[Dieball et al.(2002)]{dieball02} 
Dieball, A., M{\"u}ller, H., \& Grebel, E.~K.\ 2002, \aap, 391, 547 

\bibitem[Dirsch et al.(2000)]{dirsch00} 
Dirsch, B., Richtler, T., Gieren, W.~P., \& Hilker, M.\ 2000, \aap, 360, 133 

\bibitem[Dolphin(2000)]{dolphin00} 
Dolphin, A.~E.\ 2000, \pasp, 112, 1383 

\bibitem[Efremov \& Elmegreen(1999)]{efremov99} 
Efremov, Y.~N., \& Elmegreen, B.~G.\ 1999, New Views of the Magellanic Clouds, 190, 422 

\bibitem[Elson, Fall \& Freeman (1987)]{eff87} 
Elson, R.~A.~W., Fall, S.~M., \& Freeman, K.~C.\ 1987, \apj, 323, 54 

\bibitem[Fischer et al.(1998)]{fischer98} Fischer, P., Pryor, C., 
Murray, S., Mateo, M., \& Richtler, T.\ 1998, \aj, 115, 592 

\bibitem[Geha et al.(1998)]{geha98} 
Geha, M.~C., et al.\ 1998,  \aj, 115, 1045 

\bibitem[Girardi et al.(2002)]{girardi02} 
Girardi, L., Bertelli, G., Bressan, A., Chiosi, C., Groenewegen, M.~A.~T., Marigo, P., Salasnich, B., \& Weiss, A.\ 2002, \aap, 391, 195 

\bibitem[Gouliermis et al. (2004)]{gouliermis04}
Gouliermis, D., Keller, S.~C., Kontizas, M., Kontizas, E., \&
Bellas-Velidis, I.\ 2004, \aap, 416, 137

\bibitem[Gouliermis et al.(2009)]{gouliermis09} 
Gouliermis, D.~A., de Grijs, R., \& Xin, Y.\ 2009, \apj, 692, 1678 (Paper~I)

\bibitem[Gouliermis et al.(2006)]{gouliermis06} 
Gouliermis, D., Brandner, W., \& Henning, T.\ 2006, \apj, 641, 838 

\bibitem[Hambly et al.(2001)]{hambly01} 
Hambly, N.~C., et al.\  2001, \mnras, 326, 1279 

\bibitem[Holtzman et al.(1999)]{holtzman99} 
Holtzman, J.~A., et  al.\ 1999, \aj, 118, 2262 

\bibitem[Kerber et al.(2002)]{kerber02} 
Kerber, L.~O., Santiago, B.~X., Castro, R., \& Valls-Gabaud, D.\ 2002, \aap, 390, 121 

\bibitem[Kerber \& Santiago(2006)]{kerber06} 
Kerber, L.~O., \& Santiago, B.~X.\ 2006, \aap, 452, 155 

\bibitem[King et al.(1995)]{king95}
King, I. R., Sosin, C., \& Cool, A. M.\ 1995, \apjl, 452, L33

\bibitem[King(1962)]{king62} King, I.\ 1962, \aj, 67, 471 

\bibitem[Kontizas et al.(1998)]{kontizas98} 
Kontizas, M., Hatzidimitriou, D., Bellas-Velidis, I., 
Gouliermis, D., Kontizas, E., \& Cannon, R.~D.\ 1998, \aap, 336, 503 

\bibitem[Kumar et al.(2008)]{kumar08} 
Kumar, B., Sagar, R., \& Melnick, J.\ 2008, \mnras, 386, 1380 

\bibitem[Lightman \& Shapiro(1978)]{lightman78} 
Lightman, A.~P., \& Shapiro, S.~L.\ 1978, Reviews of Modern Physics, 50, 437 

\bibitem[Mackey \& Gilmore(2003)]{mackey03} 
Mackey, A.~D., \& Gilmore, G.~F.\ 2003, \mnras, 340, 85

\bibitem[Mackey et al.(2006)]{mackey06}
Mackey, A.~D., Payne, M.~J., \& Gilmore, G.~F.\ 2006, \mnras, 369, 921

\bibitem[Markwardt(2008)]{markwardt08} 
Markwardt, C. B. 2008, in Astronomical Data Analysis Software 
and Systems XVIII, ASP Conference Series, eds. D. Bohlender, 
P. Dowler \& D. Durand (Astronomical Society of the Pacific: 
San Francisco), in press (arXiv:0902.2850v1)

\bibitem[McKibben Nail \& Shapley(1953)]{mckibbennail53} 
McKibben Nail, V., \& Shapley, H.\ 1953, Proceedings of the National Academy of Science, 39, 358 

\bibitem[Meaburn(1980)]{meaburn80} 
Meaburn, J.\ 1980, \mnras, 192, 365 

\bibitem[Meylan \& Heggie(1997)]{meylan97} 
Meylan, G., \& Heggie, D.~C.\ 1997, \aapr, 8, 1 

\bibitem[Murray \& Lin(1996)]{murray96} 
Murray, S.~D., \& Lin, D.~N.~C.\ 1996, \apj, 467, 728 

\bibitem[Nota et al.(2006)]{nota06} 
Nota, A., et al.\ 2006, \apjl, 640, L29 

\bibitem[Gouliermis et al.(2007)]{gouliermis07} 
Gouliermis, D.~A., Henning, T., Brandner, W., Dolphin, A.~E., Rosa, M., \& Brandl, B.\ 2007, \apjl, 665, L27 

\bibitem[Oey et al.(2008)]{oey08} 
Oey, M.~S., King, N.~L., Parker, J.~W., \& Lamb, J.~B.\ 2008, IAU Symposium, 246, 65 

\bibitem[Portegies Zwart \& Rusli(2007)]{portegieszwart07} 
Portegies Zwart, S.~F., \& Rusli, S.~P.\ 2007, \mnras, 374, 931 

\bibitem[Rochau et al.(2007)]{rochau07}
Rochau, B., Gouliermis, D.~A., Brandner, W., Dolphin, A.~E., \& Henning,
T.\ 2007, \apj, 664, 322

\bibitem[Santiago et al.(2001)]{santiago01} 
Santiago, B., Beaulieu, S., Johnson, R., \& Gilmore, G.~F.\ 2001, \aap, 369, 74 

\bibitem[Schaefer(2008)]{schaefer08} 
Schaefer, B.~E.\ 2008, \aj, 135, 112 

\bibitem[Shapley(1951)]{shapley51} 
Shapley, H. 1951, Publ. Michigan Obs., 10,  79

\bibitem[Sirianni et al.(2002)]{sirianni02} Sirianni, M., Nota, 
A., De Marchi, G., Leitherer, C., \& Clampin, M.\ 2002, \apj, 579, 275 

\bibitem[Sirianni et al.(2005)]{sirianni05} 
Sirianni, M., et al.\  2005, \pasp, 117, 1049 

\bibitem[Smecker-Hane et al.(2002)]{smeckerhane02} 
Smecker-Hane, T.~A., Cole, A.~A., Gallagher, J.~S., III, \& Stetson, P.~B.\ 2002, \apj, 566, 239 

\bibitem[Spitzer(1987)]{spitzer87} 
Spitzer, L.\ 1987, Dynamical Evolution of Globular Clusters. 
Princeton University Press (Princeton, NJ). ISBN-13 9780691084602 

\bibitem[Spitzer(1958)]{spitzer58} 
Spitzer, L., Jr.\ 1958, \apj, 127, 17

\bibitem[Stolte et al.(2006)]{stolte06} Stolte, A., Brandner, 
W., Brandl, B., \& Zinnecker, H.\ 2006, \aj, 132, 253 

\bibitem[Subramaniam et al.(1993)]{subramaniam93}
Subramaniam, A., Sagar, R., \& Bhatt, H.~C.\ 1993, \aap, 273, 100 

\bibitem[Kissler-Patig et al.(2007)]{kisslerpatig07} 
Kissler-Patig, M., et al.\ 2007, ``12 Questions on Star 
and Massive Star Cluster Formation'', The Messenger, 129, 69 

\bibitem[Vesperini et al.(2008)]{vesperini08} 
Vesperini, E., 
McMillan, S., \& Portegies Zwart, S.\ 2008, IAU Symposium, 246, 181 

\bibitem[Westerlund(1997)]{westerlund97} 
Westerlund, B.~E.\ 1997, The Magellanic Clouds, Cambridge Astrophysics Series, 29  

\bibitem[Xin et al.(2008)]{xin08}
Xin, Y., Deng, L., de~Grijs, R., Mackey, A.~D., \& Han Z. 2008, MNRAS, 384, 410

\end{thebibliography}
\end{document}